# Balancing India's 2030 Electricity Grid Needs Management of Time Granularity and Uncertainty: Insights from a Parametric Model

Rahul Tongia, Ph.D., FNAE
Senior Fellow
Centre for Social and Economic Progress, 6 Rizal Marg, New Delhi, India
rahul.tongia@csep.org
ORCID: 0000-0001-6808-7534

## 1. Abstract

With some of the world's most ambitious renewable energy (RE) growth targets, especially when normalized for scale, India aims more than quadrupling wind and solar by 2030. Simultaneously, coal dominates the electricity grid, providing roughly three-quarters of electricity today. We present results from the first of a kind model to handle high uncertainty, which uses parametric analysis instead of stochastic analysis for grid balancing based on economic despatch through 2030, covering 30-minute resolution granularity at a national level. The model assumes a range of growing demand, supply options, prices, and other uncertain inputs. It calculates the lowest cost portfolio across a spectrum of parametric uncertainty. We apply simplifications to handle the intersection of capacity planning with optimized despatch. Our results indicate that very high RE scenarios are cost-effective, even if a measurable fraction would be surplus and thus discarded ("curtailed"). We find that high RE without storage as well as existing slack in coal- and gas-powered capacity are insufficient to meet rising demand on a real-time basis, especially adding time-of-day balancing. Storage technologies prove valuable but remain expensive compared to the 2019 portfolio mix, due to issues of duty cycling like seasonal variability, not merely inherent high capital costs. However, examining alternatives to batteries for future growth finds all solutions for peaking power are even more expensive. For balancing at peak times, a smarter grid that applies demand response may be cost-effective. We also find the need for more sophisticated modelling with higher stochasticity across annual timeframes (especially year on year changes in wind output, rainfall, and demand) along with uncertainty on supply and load profiles (shapes).

## 2. Keywords

Electricity grid, modelling, capacity expansion, economic despatch, uncertainty, renewable energy

## 3. Introduction

The modern alternating current (AC) grid has been described as the greatest engineering achievement of the 20[th] century (US National Academy of Engineering 2021). One of its key features is that it continuously requires real-time balancing, except to the extent one may have storage. In fact, one requires more supply capabilities than instantaneous demand to allow for a system buffer. Technical specifications for buffer requirements and spinning reserves (meant to be 5%) are specified in the Indian Electricity Grid Code (IEGC) and affiliated notifications.[1] Another key facet of this system is the interconnectedness across hundreds of thousands of nodes for grid like India, or hundreds of millions if end-users are included. There is a shift in power flows away from traditional centralized models, with more generation occurring at the edge such as from rooftop solar. However, the majority of power flows is still from larger generators or at least sources separate from the end-user, and this is the focus of the paper.

---

[1] The specified 5% spinning reserve, based on Section 5.2.3 of the National Electricity Policy (Ministry of Power, 2005) at the national level, isn't always met, and remains an area of focused effort (CERC 2015).

Power system planning and optimization is a complex engineering task overlaid with economic constraints to attempt to determine the optimal supply (fuel) mix. This covers issues of supply adequacy, least cost, and other constraints including environmental, energy security etc. There are additional complications based on path dependencies, issues of timescale, investment, etc., which are distinct from uncertainty in demand and supply conditions. The uncertainty may be stochastic, and there also remains uncertainties over price points, especially for new technologies. There are further complications based on fuel supply, transmission congestion, etc. Hobbs (1995) summarizes the evolution of basic planning, which evolved from Linear Programming to Mixed Integer Linear Programming and added uncertainty, reliability, etc. In addition to purely technical analysis, market design issues lead to further choices and techno-economic coupling. An updated framework for planning and markets under high RE is by Lopes and Coelho (2018).

## 3.1. Motivation and Objectives

This paper presents a granular analysis of balancing India's electricity grid 2021 through 2030 at a national level with a half hourly resolution. By applying parametric analysis under uncertainty, this combines capacity expansion with economic despatch analysis across fuel choices (based on available capacities).

There are a number of other studies on India's grid as well as worldwide that focus on technical issues, detailed below. Our focus is less on computational techniques and modelling per se, but rather understanding the ranges of uncertainty and the factors that matter from a technical and economic standpoint, as well as trendlines and cross-over points.

Some of the specific questions that we investigate include the below:

a) How will India manage its extremely aggressive renewable energy (RE) plans?[2] How much RE is feasible, and are there risks of "too much" RE, which would lead to curtailment?
b) Does India need more coal-based power? Does it need to build more coal power plants?
c) What supply (fuel) options make the most sense for new capacity going forward? This isn't just a function of their economics but also issues like system security and managing volatility or uncertainty.
d) Are batteries the answer in the short term or longer-term? How should they be thought of for planning?
e) What are the factors that matter for grid planning and policy?

Another contribution of this paper is the demonstration of analysing uncertainty by using simplified economic despatch models across wide parametric (parallel) uncertainty.

## 3.2. Literature review and other studies

While there have been a number of studies on India's grid in recent years, they differ in terms of their methodology and objectives.

Power system modelling has improved and evolved over the years thanks to increases in computational capabilities. The number of nodes that can be handled for load flow analysis has increased in and the time resolution has not only approached virtually real-time, but there is also extensive work towards predictive models and real-time state estimation, especially driven by the use of synchrophasor measurement units (PMUs) (Chandra *et al.* 2016). Adding a layer of choices for capacity expansion planning adds multi-fold possibilities for optimal system planning, to the extent that most practical models need specialized techniques for coupled expansion planning with optimal supply mix. The latter focuses on how best to operate a system *once built*, but doesn't directly tell us what is the best system to build. Table 1 shows a comparison between different modelling and computational techniques.

---

[2] India aims to more than quadruple its traditional RE (excluding hydro) between 2021 and 2030, from a little over 100 GW to roughly 450 GW. Adding in hydro gives over 500 GW of total RE, in sync with Prime Minister Modi's Glasgow COP26 declarations (PIB 2021b), which, technically, were for "non-fossil", i.e., including nuclear power).

**Table 1 Coverage of reliability by different types of models**

| Model Type | Generator (Resource) Adequacy | Flexibility Requirement | Transmission Adequacy | Generator Contingencies | Transmission Contingencies | Frequency Stability | Voltage Stability, Voltage Control |
|---|---|---|---|---|---|---|---|
| Capacity Expansion | Often | Somewhat/Depends | Typically No | No | No | No | No |
| Unit Commitment and Despatch (Production Cost) | Yes | Yes | Partially | Somewhat | Somewhat | Somewhat | No |
| Network Reliability (AC Power Flow, Dynamics) | No | No | Yes | Yes | Yes | Yes | Yes |

Source: Boyd (2016)

The simplest models aggregate the entire grid to the geographic area of interest, e.g., national or state. At the extreme, these can ignore transmission. Most models rarely compute very high time resolutions, which focus on sub-cycle grid stability and transients.

If we look at the list of recent Indian studies for the grid, almost all of them are focused on determining the optimal fuel mix, and none of them focus on narrow engineering issues like transients. Several of them do examine transmission issues but aggregated at the state level (each state operates like a node). There are other constraints that are either difficult to model or purposely left out due to choices made either for simplification or for segregating political from technical constraints. These can include issues of fuel supply, contracting (including asymmetric contracts that segregate between fixed and variable, i.e., fuel costs), etc.

Even with "perfect" or infinite transmission, the intersection of capacity expansion and optimal production once built is highly complex, in part because different fuels or supply options vary in their split between fixed and variable costs. Helistö *et al*. (2019) show the increasing sophistication possible in modelling (Fig. 1).

**Fig. 1 Classification of modelling approaches**

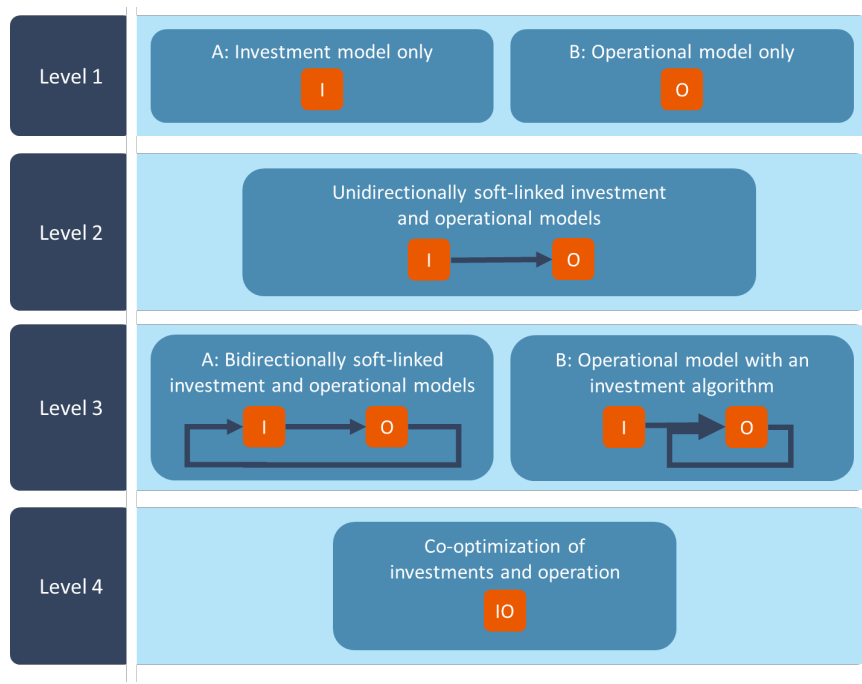

*\* Source: Helistö et al. (2019)*

The analysis in this paper can be considered level 3A, based on a feedback loop from optimal operations to investment choices (Level 3). This is not via an optimization model, but parametrically over a wide n-dimensional space, i.e., in parallel, to handle a wide range and allowing pseudo-optimal through lowest cost (or other criteria). Thus, subject to the limitations like national despatch at an aggregate level, without transmission constraints or unit level constraints, this analysis still provides insights approaching co-optimization models. A national model is still relevant as it avoids limitations of state-level studies which cannot properly capture the incentives or dynamics of out-of-state supply, except, typically, amounts contracted for through Power Purchase Agreements, such as from Central Generating Stations (CGS).

One of the general challenges for all modelling exercises is underlying data, both availability and quality. Timeseries data is especially difficult but important for a system that is both evolving as well as one that has natural variability and stochasticity. Techniques used by different researchers include using limited data sets, synthetic data sets, or estimations based on secondary data including satellite data. Many studies use wind and solar profiles as per alternative models (e.g., Palchak D and Cochran J (2017) used satellite data for the NREL Greening the Grid study, first undertaken for all-India 2022), instead of actual data of wind or solar-based generation. Tongia, Harish, and Walawalkar (2018) have found enormous discrepancies in RE output based on actual production compared to satellite model-based data like with NREL.

Even if the aggregate (average) discrepancy between modelled data like wind output versus actual output is small, the instantaneous discrepancy can have an enormous impact on grid design. This is because the aggregate deals with total energy, while instantaneous output, which is important for real-time balancing. Surplus of 10% in one time period cannot compensate for a 10% deficit in a different one, even if the total is the same (and and total energy or prices are the same).

This study does not limit itself to time-sampled or representative days, which has been used in a number of studies due to data and/or computational limitations. Mallapragada *et. al.* (2018) find that the resolution is a key issue for comparing sampled (both time-slice and chronological Capacity Expansion Models) to full dataset production cost modelling.

Table 2 shows a summary of recent studies for India. The high ranges of capacity shown for some fuel mixes portend practical challenges in growth, e.g., almost doubling hydro in 9 years (e.g., BNEF and TERI). Nuclear power is also unlikely to materialize at the high capacity shown in BNEF's analysis, as we can monitor already planned or under-construction plants.

**Table 2 India Installed capacity (GW) – 2020 (actual) and 2030 projections from recent studies**

| Technology | Actual (2020) | CEA (2030) | NREL (2030) | BNEF (2030) | TERI (2030) | LBL (2030) Primary Least Cost |
|---|---|---|---|---|---|---|
| Coal | 206 | 267 | 170 | 234 | 238 | 229 |
| Natural Gas | 25 | 25 | 49 | 25 | 25 | 25 |
| Nuclear | 7 | 19 | 11 | 33 | 17 | 19 |
| Hydro | 54 | 61 | 54 | 81 | 84 | 62 |
| Wind | 38 | 140 | 200 | 109 | 169 | 142 |
| Solar | 36 | 280 | 250 | 204[a] | 229 | 307 |
| Battery storage | 0 | 27 (4-hour) | 16 (2-hour) 68 (4-hour) | #N/A | 60 (2-hour) | 15 |
| New pumped storage | N/A | 10 | 1.5 | 0 | 0 | 63 (252 GWh)[b] |
| Load shifting | | 0 | 0 | 0 | 0 | 60 |
| Other RE | 15 | 15 | | | | |
| Total | 381 | 844 | 824 | 734 | 822 | 862 |

*Source: Abhyankar et al. (2021). The first four 2030 studies are as summarized (verbatim); the last column summarizes Abhyankar et al.'s findings.*

[a] *AC capacity with an Inverter Loading Ratio of 1.30, implying the DC capacity to be approximately 265 GW.*

[b] *Modelled as standalone battery energy storage systems.*

\* *Each study uses a different set of assumptions on technology costs, baseline year, and operational parameters. Therefore, the comparison across studies is shown for illustrative purposes only and should be interpreted carefully.*

Our analysis does not compute the optimal fuel mix, but rather finds costs across a range of mixes, and the lowest cost can be determined based on the cost frontiers.

## 4. Methods and Objectives

Table 3 summarises the model, which balances demand with supply through iterative steps, starting with simple despatch but then adding sophistication such as limits on coal power plant operations and secondary displacement of fossil fuels using new capacity added over time. A key output from this phased approach is to examine if existing capacity is sufficient to meet demand over time, inclusive of planned expansion of RE and modest growth of hydro and nuclear. In case supply is insufficient, this leaves a residual to be met with something NEW[3]) at lowest costs per time period, for each of the successive modules or steps. Most variables such as RE capacity are varied parametrically (in parallel) across a wide range. More details on the steps below are given subsequently.

**Table 3 Model Steps for Optimal Economic Despatch**

| | Model Step | Outputs |
|---|---|---|
| 1 | *Produce shape curves through 2030 – demand and RE*<br>Scale demand and installed capacities of fuels available through 2030 (pro-rata demand and RE shapes) [a] | 1. Demand over time<br>2. Fuel-wise net capacity over time |
| 2 | *Calculate modified "net demand"*<br>Subtract "must run" or zero marginal cost supplies (RE, hydro, nuclear) from demand giving modified net demand | 1. Modified net demand (to be met by fossil fuels)<br>2. Interim curtailment of RE (based simply on demand and must run) |
| 3 | *Find optimal supply mix based on capacity assumptions*<br>Run Economic Despatch for the modified net demand with fossil fuels [b] | 1. Fuel-wise supply per time period<br>2. Unmet (residual) demand |
| 4 | *Iterate to add coal power plant flexible operations (part load) constraints*<br>Calculate limitations on coal output (daily basis) for "flex" (flexible, or part load) operations [c] | 1. Updated fuel-wise supply per time period (inclusive of coal flex limits)<br>2. Updated RE curtailment (inclusive of coal flex limits) – pre-storage<br>3. Unmet demand requiring NEW supply<br>4. Economics of existing supply and capital investments for growing RE |
| 5 | *Estimate costs (capex, fuel and O&M) for NEW supply to meet residual demand – sized for capacity requirements WITH feedback loops using NEW capacity to displace existing fossil fuels*<br>Compare options for NEW supply (to meet otherwise unmet demand) via new capacity added annually [d] | 1. Capital, fuel, and O&M costs of NEW<br>2. Impact of NEW on existing fuel mix and corresponding economics ("feedback loops")<br>3. Fuel Supply mix after NEW plus any feedback loops |

---

[3] NEW in capitals emphasises this is the category of supply for otherwise unmet demand, and not just an adjective.

| | | |
|---|---|---|
| 5A | Add an option to undersize batteries or coal, with residual unmet demand to be met via biodiesel | |
| 5B | If NEW is coal, calculate displacement of gas (feedback loop) [c] | |
| 5C | If NEW is batteries, calculate requirements for dedicated RE to charge the battery, both minimum to meet NEW and maximum to fully utilize the battery daily (beyond NEW requirements) [f] | 1. Estimate additional RE (solar) required to charge the battery if curtailed RE is insufficient to charge the battery. |
| 5D | If NEW is batteries, calculate displacement of gas and coal (feedback loop) | |
| 5E | If batteries displace coal, calculate the bonus based on reducing daily peak coal supply, which reduces daily RE curtailment due to flex limit shifts [g] | |
| 6 | *Find optimal mix based on lowest cost* Calculate NPV (net present value) of different RE levels and NEW options (including under-sizing, adding dedicated solar for the battery, etc.) [h] | 1. System level NPV – usage of existing fleet, planned RE, and NEW supply (excludes sunk costs and costs of nuclear/hydro = must run or zero marginal cost) 2. Levelised per unit (per kWh) costs of existing capacity plus planned RE (excluding sunk costs) 3. Levelised per unit (per kWh) costs of NEW supply including feedback loops |

[a] *Assumes modest growth of hydro and nuclear, some retirement of existing coal capacity. RE grows measurably as variable RE (VRE), projected across different ratios of solar and wind.*

[b] *Coal and gas are each segregated between capacity that outputted in 2019 vs. not used in 2019 per time period. The latter is underutilized (slack) capacity.*

[c] *For each daily output of coal, examine maximum coal (MW), and apply flex limits (base of 60%) for the minimum operations of coal (MW) allowed for that day. If allowed minimum is greater than level by despatch, the constrained minimum is applied, which leads to further curtailment of RE (and higher coal).*

[d] *Based on estimates for capital and fuel costs.*

[e] *The fuel economics mean, in practice, NEW coal would displace more expensive non-APM natural gas; same for batteries. We do not spread coal prices across plants except for 2019 run coal (a fraction of the fleet) assumed to be less expensive than if all the capacity ran. Thus, we do not estimate use of NEW coal to displace old coal.*

[f] *Additional RE (solar) is required if curtailed RE is insufficient for charging the battery. Beyond the minimum solar for the battery to meet NEW supply, additional solar can harness undertilised battery capacity.*

[g] *The battery directly displaces coal (Step 5d) but also has a secondary or bonus impact on lowering RE curtailment by lowering peak coal. SI 2 has more details on the methodology for feedback loops of batteries and coal.*

[h] *The costs of NEW include benefits of feedback loops, which lower fossil fuel outputs from Step 4. Economic benefits of lower gas or coal due to a NEW option like a battery are allocated to the battery.*

## 4.1. Curves for time varying supply and demand

This study is a half-hour resolution economic despatch grid balancing study for India between 2021 and 2030, operating at the national level. The analysis is limited to utility grid operations, and thus excludes both captive power (directly owned by end-users) or behind the meter generation technologies like rooftop solar.

A significant base data set comes from carbontracker.in (CSEP Electricity and Carbon Tracker, 2021), a portal developed by the authors' team that gathers data through scraping from publicly available resources such as MERIT (Ministry of Power, 2021), by the government of India. The data covers the instantaneous load met[4] (or grid level demand), as well as the supply by fuel all-India. We use 2019 data as being the most recent full calendar year before Covid.[5] For computational reasons, we limit ourselves to half hourly data points. We separately have data from the Central Electricity Authority (CEA) indicating capacities by fuel as well as other operating parameters such as auxiliary consumption (the portion of the gross generation consumed in-plant and thus not available to the grid as net generation).

Time varying data are a key component for the base of 2021 (starting point of the analysis), which are then escalated at various projected rates through 2030. For example, overall electricity demand could grow anywhere, say, between 5% to 5.5% per annum on average. For the base of 5.25% growth, this translates to 2160 billion kWh or billion units (BU) of demand in 2030. Meeting such demand at a state boundary level requires a little under 10% more generation to cover auxiliary consumption (which is falling with the rise in RE) and inter-state transmission losses.

In this analysis, except where stated explicitly, growth is assumed proportional or pro rata, and we do not change the load shapes.[6] Key data curves with time of day (ToD) implications include: (1) demand shape; (2) RE output, which will vary with shares of solar and wind; (3) supply by each respective fuel in 2019. Table 4 summarizes initial data for *maximum net* (post-auxiliary consumption, post minimum maintenance schedules) available supply by fuel type.

**Table 4 Expected 2021 Expected *net* fleet-wide capacity by fuel (based on 2019 performance and growth)**

|  | Maximum Busbar Capacity Available [a] (GW) |
|---|---|
| **RE** | 98.0 |
| **Hydropower** | 35.5 |
| **Natural Gas** | 21.3 |
| **Nuclear** | 5.4 |
| **Coal** | 162.6 |
| *Total* | 332.7 |

[a] *This capacity differs from installed capacity, which is typically higher, because the latter is gross capacity, i.e., before removing auxiliary consumption. Some fraction of the fleet is also expected to be unavailable due to periodic maintenance. The practical available capacity would vary over time, especially for RE.*

This available capacity of many fuels such as gas and coal in 2019 was higher than they actually delivered in most time periods throughout the year. For despatchable fuels, this represents slack capacity that could be drawn in subsequent years as per the despatch modelling exercise. In the base case, we assume no net addition of coal or gas capacity, rather, parametrically assume a small fraction of the existing coal fleet will retire by 2030. We also assume the addition of pollution control equipment such as flue gas desulfurization (FGD), which imposes an operating efficiency penalty of 2.5% (relative basis) as the technology is deployed.

---

[4] "Load met" in the Indian context is a grid level supply measure that is after busbar (generation station post auxiliary consumption) and at the peripheries of state boundaries. It is before accounting for transmission and distribution losses, and, hence, not a measure of end-user consumption.

[5] We apply simple heuristic error corrections to the data, which have occasional freezes or missing data. The data are also significantly underreporting real-time RE, probably due to limited instrumentation such as through SCADA. We apply a proportional correction factor to match aggregate RE output.

[6] While some amount of load shape changes is organic and evolutionary, there are specific policy initiatives to try and shift the load to match supply such as the solarisation of irrigation pumpsets. Abhyankar *et al.* (2021) deploy load shifting as a mechanism for grid balancing. While load shifting does benefit raising demand during increased solar supply, policies such as solar irrigation pumpsets are unlikely to reduce the evening peak demand when solar supply dies down. This is because most states do not supply agricultural load during the evening peak, which is driven by residential and commercial demand. They typically provide 6-8 hours of agricultural supply in a day, rostered or controlled at a phase or feeder level. Thus, such load shifting can help reduce curtailment mid-day, but not proportionally reduce the need for NEW in the evening.

As demand grows over time, it can be met through a combination of (1) slack capacity; (2) planned capacity expansion across RE, hydro or nuclear, or (3) "something new" built, which could be a range of fuel choices including coal, natural gas (combined cycle, open cycle, spark combustion, aka CCGT, OCGT, and gas IC), diesel/biodiesel, and batteries. These span a range of capital-cost intensive to fuel-cost intensive solutions. The model first calculates the ability to meet demand using slack and planned expansion RE/hydro/nuclear, and then outputs a residual of unmet demand which is subsequently reanalysed for lowest cost service with a new build.

Planned growth of hydro and nuclear is conservatively projected to be modest, at 3% and 3.9% annually, respectively.

The most important and largest growth is expected to come from RE, which is part of the government's ambitious plans announced even before pledges made at COP26 in Glasgow under the UN Framework Convention on Climate Change (PIB 2021b). The base analysis treats RE as variable RE (VRE), i.e., RE without storage, which is treated separately (as lithium or comparable batteries). One reason for this split is to understand the potential and limits of VRE, which is the lowest cost form of RE deployment.

To model uncertainty, we parametrically vary the total of RE between 250 to 550 GW. We assume a range of ratios between solar and wind, which is important given their output characteristics (diurnal versus seasonal). We also assume forward-looking capacity utilization factors, or CUFs (aka Plant Load Factors, or PLFs) for solar and wind, which are higher than present averages.[7]

An important form of uncertainty is the year-on-year RE expansion increment to achieve a 2030 target. Is, say, 270 GW of growth over 9 years 30 GW per year, linearly? Instead, we assume a continuous growth, leading to a compound annual growth rate, instead of a linear rise. In reality, all variables including demand might have annual ups and downs. Given the high levels of uncertainty across technical, operating, and economic parameters, the focus of this study is less on any specific number (such as tons of fuel or rupees per kilowatt hour) but rather on trends, crossover points, and determining the key factors that matter.

## 4.2. Despatch by fuel and price

Based on the net capacities available, the model begins with a first round of half hourly economic despatch using a modified-merit order by fuel type to determine lowest cost operations.[8] This is not a full merit order despatch as the grid operator would do because it is not undertaken at the plant level, rather limited to the aggregate fuel level with some heterogeneity listed below.

Importantly, we apply time-block-wise despatch analysis only upon remaining demand that is not met by supply options that are either must run or zero marginal cost. Thus, the demand for a particular time period is first partly met by RE under starting assumptions of minimized curtailment as well as nuclear and hydro following proportional shapes as 2019.[9] The remaining demand is met via existing coal supply as used in 2019, existing

---

[7] Higher PLFs are based on updated engineering designs for solar with higher DC to AC oversizing, and higher hub heights for wind at superior locations. These can be as high as 27% and 35% for solar and wind, respectively, for new builds. Given the 2019 base split of solar and wind isn't available, we estimate the output of wind based on subtracting solar from the RE output, to find a ToD shape per MW. The solar estimate is calculated using NREL's PVWatts calculator (NREL, 2021) for a representative north Karnataka location, which would crudely approximate the national blend. More precision is difficult and likely unnecessary given annual variations likely, and left for a future stochastic analysis. Given the 2019 data were for wind with a lower PLF than projected in the future, the wind shape-curve estimated from 2019 is scaled to achieve 35% PLF for future wind to produce the per MW wind curve by ToD. Future solar is also modified from historical to achieve not just a higher PLF but, specifically, a flatter-at-the-peak output during the middle of the day based on DC to AC oversizing.
[8] The analysis is undertaken in the Analytica modelling environment, by Lumina Systems, which specializes in analysis under uncertainty.
[9] While hydro has despatch flexibility, its total daily availability is heavily constrained by the monsoon, which determines annual output, as well as competing duty cycle uses for water flows such as irrigation. Sengupta (2021) shows how hydro doesn't follow normal economic despatch based on marginal costs based on a reduced form all-India state-wise despatch model, and instead operates like a peaker. Given our demand curves are

gas used in 2019, underutilized coal (slack), and then underutilized gas (slack). Such despatch either leads to meeting demand or a residual unmet demand which has to be met by something new.

Segregating coal and gas between 2019 usage versus underutilized (slack) capacity allows them to be priced at different fuel costs. We assume that 2019 operations roughly followed economic despatch which gives us a prioritization order of costs between coal and gas of 2019. For slack capacity, we assume coal is less expensive than gas except for 2019 gas, which found use in 2019 and is thus assumed to be less expensive than future (slack) gas capacity. As more and more coal fleet becomes required, it will be plants that are more expensive at the margin, typically located at further distances from the coal mines. However, for simplification purposes, the marginal costs of slack coal are all treated as a group, instead of differentiating at a plant level. We assume no constraints on the availability of fuel. We apply a wide differential on gas prices between 2019 gas (assumed to be the cheaper Administered Pricing Mechanism, or APM, gas) and slack capacity gas.

Importantly, this is not a unit commitment analysis (which tells us which plants should be available or turned on). For this reason, there is no internalized constraint based on plant start or stop costs. Instead, we minimize daily start stops based on an assumption of flexible (part load) operations for coal plants can parametrically be between 55% - 70% and apply this criterion daily. National guidelines specify a technical minimum of 55% (CERC 2017) but a large number of plants, especially older plants, appear unable to comply with such flexibility. Hence, our fleetwide average has a baseline of 60% part load capabilities per day, which is calculated based on the maximum coal supply per day. We also do not hardwire ramping constraints as an input to the model but rather validate the results post facto to check that aggregate ramping remains within fuel type capabilities. A conservative requirement would be 0.5% per minute for coal, even though CEA expects 1% per minute ramping capabilities (CEA 2019a)

Other technical constraints include requirement to have a capacity buffer at a fleet level, of 5% per IEGC. Details on assumptions are given in SI 1.

To meet future demand that cannot be met by existing capacity or planned growth of RE/hydro/nuclear, the model compares total costs across a wide range of fuel choices including new coal capacity, natural gas (combined cycle, open cycle, and engine), diesel (including biodiesel), and storage (battery).[10] While we do not apply full hybrid modes of capacity expansion across all the choices, we do estimate costs for the system without full battery storage that supplements energy through biodiesel, and similarly under-sizing NEW coal to be complemented by, say, biodiesel (a zero-carbon option).[11] The advantage of such under-sizing is that it avoids disproportional capital costs for an infrequent peak, even though the fuel cost of biodiesel may be very high on a per kilowatt-hour basis.

For batteries, literature such as NREL's studies by Cole (2021) often assumes a combination of capacity and duration (e.g., "4-hour battery", $200/kWh = 0.25 kW for 4 hours). In contrast, we segregate these parameters explicitly allowing the model to solve for necessary sizing, spanning both energy and capacity requirements. This treats batteries as a capital cost instead of as the equivalent of a despatchable supply (measured through levelized cost of energy, or LCOE, based on average Rs./kWh supplied). Model capacity costs (for inverter and balance of systems) per kW are kept aggressive, close to $100/kW, or 7,500 Rs./kW.[12]

---

scaled from 2019, we apply hydro's daily peaking cycle proportional to 2019. In the future, better peak pricing and signalling will help optimize hydro despatch. However, the choice to model hydro similar to 2019 doesn't affect our despatch because even if it were part of full economic despatch with optimal signalling, it would be used before other fuels given its low costs (*subject to water availability*). In our case, we utilize RE, nuclear and hydro before other (fossil) fuels in the modified-merit-order despatch.

[10] We do not distinguish between under-construction coal power plants and greenfield power plants even though the former may be less expensive at the margin, treating historical costs as sunk. However, some of the under-construction capacity may simply displace existing capacity that either retires because of end-of-life or is forced to retire due to inability to comply with upcoming pollution control norms.

[11] Hybrid options are limited based on the expected tradeoff for high and low costs for capital vs. operations (fuel).

[12] This is as based on discussions with experts at the India Energy Storage Alliance (IESA).

## 4.3. Economic Objectives

The model's primary output is system cost subject to meeting demand and other technical constraints. *Importantly, the model does not calculate full system costs but rather calculates __additional__ costs compared to reference costs that are treated as sunk.* Thus, capital costs of existing installed capacity are not part of the output space and is treated as a sunk cost.[13] Similarly, even for expansion of capacity, planned hydro and nuclear are treated as exogenous and sunk costs, more so because we do not vary their levels; their growth will be constrained by development capacity. However, given the key decision variable is the quantum and form of RE, and hence capital costs for such growth are included in the cost analysis. Similarly, fuel costs for existing capacity are also integral to the calculations, separated between 2019 output levels and slack output with different marginal costs for gas. SI 3 has more details on prices. Lastly, the model calculates the total costs for anything NEW added to meet demand, spanning capital, fuel, and O&M costs.

The primary output is the total system cost (excluding sunk costs) in net present value (NPV), which allows us to calculate the optimal growth of RE and options for NEW supply. This also means the model is not designed to calculate the average systemwide cost (Rs./kWh). A reason for not doing so is the lack of available data on costs and contracts for existing capacity. The government's MERIT database lists some variable costs but those are not necessarily reflective of what is ultimately paid by utilities at the margin.

All numbers are nominal rupees (or dollars), and we separately use a discount rate (different from the weighted cost of capital, or WACC[14]) to bring system costs into NPV terms. We use a base discount rate lower than WACC, in part because inflation is lower and also because the average rate of rise in wholesale price of electricity has been much lower than costs of capital. As an example, the base capital costs and O&M costs for solar and wind are such that there is a mild fall in per unit costs for new builds, which is a substantial decline in real terms.[15] Solar capital costs have a base annual decline in nominal rupees of 2% annually. Foreign exchange depreciation is a variable explicitly used for battery costs.[16]

## 5. Results

## 5.1. Baselining from 2019

The performance of the grid in 2019 is important to understanding its evolution through 2030 (Fig. 2). We assume after 2021 the effects of Covid are diminished and the economy mostly recovers. 2021 is treated for the model quite similar to 2019, given 2020 was a (likely temporary) blip; any downturn is assumed to recover through 2021, or by 2022.

**Fig. 2 Capacity and generation shares 2019**

---

[13] O&M costs for the existing fleet are also treated as sunk (fixed) costs, and thus not part of the NPV calculations. The small level of variable O&M is embedded in fuel costs. In practice, these will vary significantly based on duty cycle, but that is beyond the scope of this analysis and likely to be second-order.

[14] The likely weighted average costs of capital (WACC) across fuel choices for new growth will vary, with green technologies finding lower interest rate loans. However, we use a mid-level WACC equally across technologies to emphasize the importance of shifting interest rates in finding crossovers between choices. 8.5% is the base rate reflects equity returns close to 13%, and lower rate debt. Lowering the WACC within bounds as used lowers all costs, but doesn't change the cross-overs.

[15] The base wind prices and solar prices are slightly higher than record bids, such as Rs. 1.99/kWh for solar (PIB 2021a) not only to be conservative but because the very high volumes anticipated would require pricing commensurate with "non-winning" bid prices of today.

[16] Domestic manufacturing may increase with the government's Production Linked Incentives (PLI) scheme, but some raw materials will remain imported, and, in the short run, savings from PLI might only bring parity with global pricing, and not necessarily beat these.

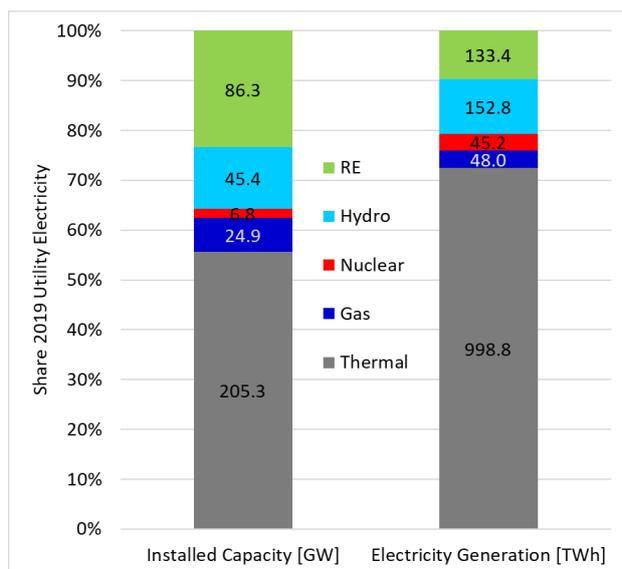

*These are calculated from CEA (2019b) data, including various Monthly Executive Summary reports.*

The supply mix of 2021 (Fig. 3) has two main characteristics of importance. First, different fuel options have different characteristics over the day. RE has zero marginal cost and thus whenever available, used in full (subject to limited technical curtailment). Hydropower cycles up and down to meet demand, which typically has a bimodal distribution (morning and evening peaks). Second, there is enormous slack in the system based on underutilized capacity of coal and gas.

**Fig. 3 Chronological fuel mix 2021**

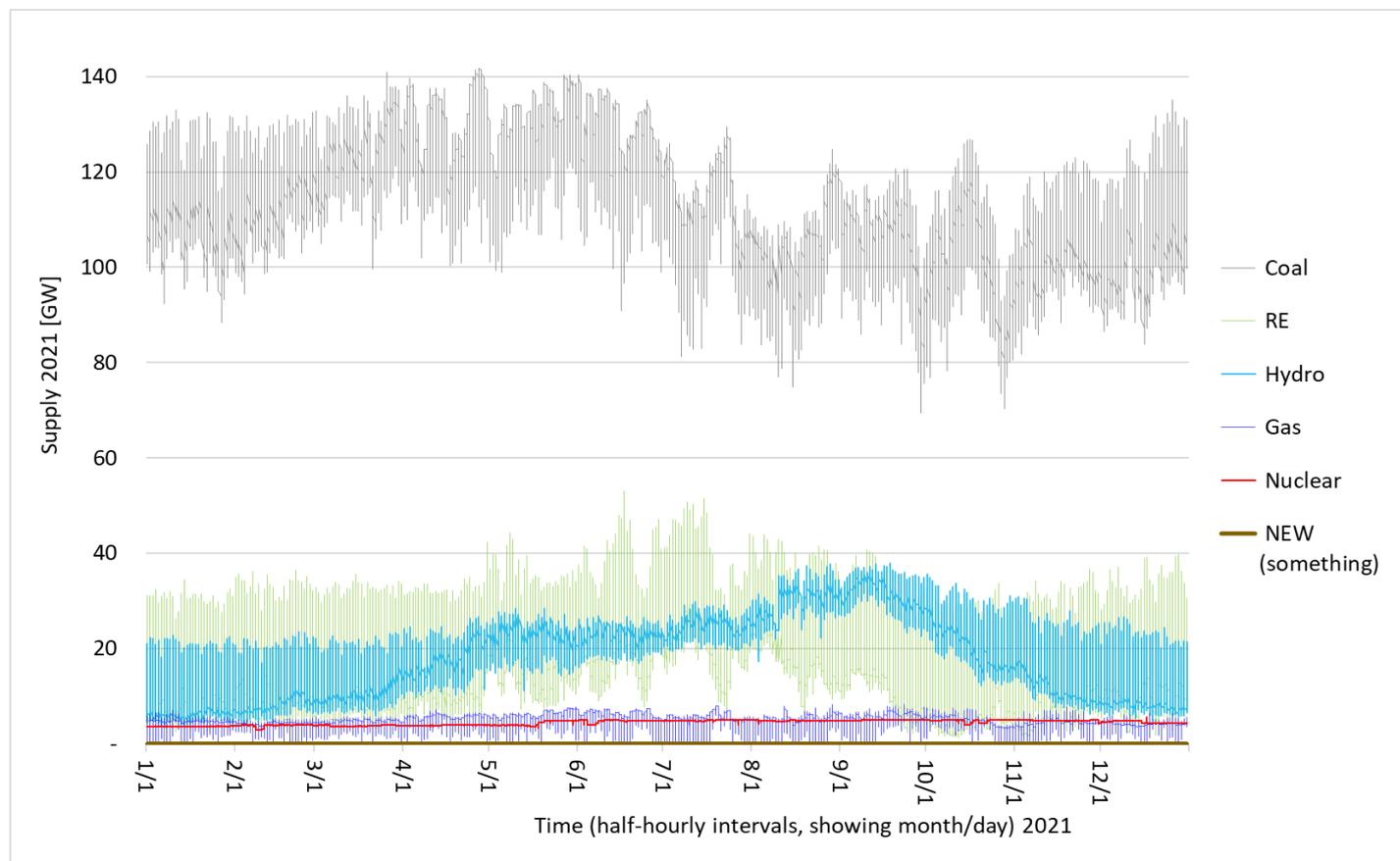

*\* This shows the net supply mix at a busbar level, post auxiliary (in-plant) consumption. In contrast, costs per unit are typically calculated as gross, prior to auxiliary consumption; this also raises the required gross generation. Zero "NEW" means capacity is sufficient to meet demand.*

While renewable energy, especially solar, coincides with much of the demand, the evening peak does show a period of high ramping, especially when measured on a "net demand" basis (remaining demand after treating RE supply as negative demand). This has been called the "duck curve", first identified through the California Independent System Operator (Jones-Albertus 2017). This emphasizes the limited contribution of RE to the evening peak, which falls to 2% or lower on selected days, an order of magnitude lower than its capacity share, and about 5 times lower than its share by energy on an annual basis. Wind especially manifests extreme seasonality. If we compare the ratio of daily max to min RE output, post monsoon season it can be as high as about 28 times.

Time-series data from CEA for demand growth shows enormous year-on-year variation. Our estimate for a longer-term average (FY2007-2020) for grid-based electricity generation has been 5.58% CAGR, but recent years have seen a slow-down, which might also manifest due to increased consumer self-generation or third-party sales, especially with RE. Thus, our base assumption is 5.25% annually.[17]

For 2021, post ISTS losses, demand starts at 1,360 billion kWh a.k.a. billion units (BU) and so a 5.25% growth indicates over 71 BU of demand rise in the short run. In contrast, historical growth of RE through FY2020 has only been BU 24.9 BU at the highest growth, much less than total demand growth. Thus, unless RE capacity growth increases, meeting demand will require more intensive utilization of existing assets in the short run. This is especially true for the projections made in the model which have an exponential growth over time for RE capacity, i.e., RE output grows faster over time.

## 5.2. Key Findings

Rising RE is a key contributor to meeting demand, but is not entirely sufficient, starting with the near term as we saw before. Coal utilization rises in an absolute sense even though its share of energy declines from 72% in 2019 (Fig. 2). Fig. 4 shows the 2030 mix based on economic despatch over a wide range of RE in 2030. Even with very high RE (450 GW), coal is still almost half the supply. Of course, this is before considering what is the source of NEW output.

**Fig. 4 Fuel Mix (economic despatch) in 2030 for varying RE capacities (without feedback from NEW supply)**

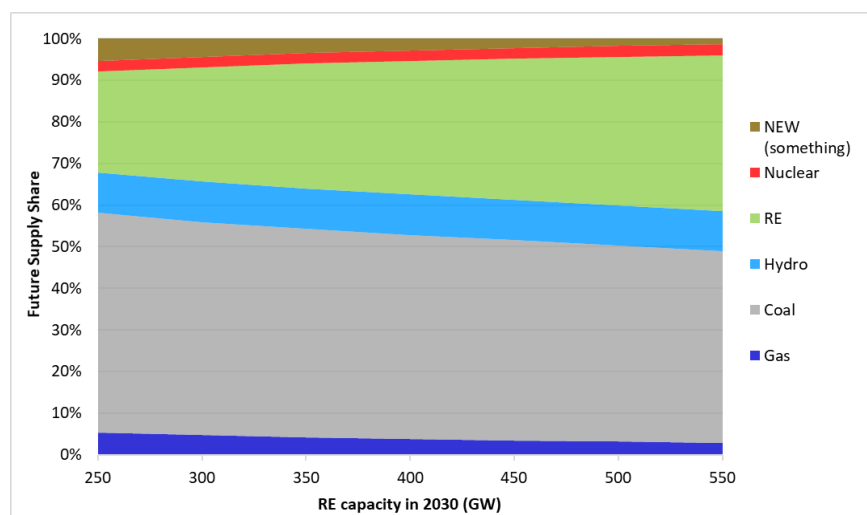

---

[17] On the other hand, there is a move to electrify more sectors across the economy, including mobility, cooking, and industry. CEA (2019c) found an electricity to GDP elasticity of 0.74%, but with their assumed GDP growth through 2030, found a demand growth close to 5%. Our estimates based on trends estimate a higher elasticity, but slightly lower GDP growth over a long time period, and hence 5.25 is a fair mid-point.

*\* This assumes 5.25% growth of demand, 2:1 solar:wind ratio, and 60% flexible operations (reduced load) limits for coal power, and also similar load profiles (time of day shapes) as 2019. Unless stated otherwise, all figures and tables are for these base parameters.*

*\*\* The RE is after accounting for curtailment. These are before feedback loops from NEW supply, e.g., NEW coal could displace expensive gas, or a battery could displace fossil fuels.*

Fig. 5 shows the chronological output for 2030, where rising demand means existing slack capacity of coal and gas are important to meet demand, especially on a time-of-day basis. This emphasizes the importance of choices over retirement of existing coal stock prematurely. Present coal surplus will not last beyond the middle of the decade, and on a capacity basis it runs out before it does on an energy basis (to meet the grid buffer technical requirement of 5%). One measure of the value of such capacity can be seen in Fig. 12 subsequently, where we see that coal and natural gas use from existing capacity as available both grow by 2030 based on economic dispatch even under a scenarios of high VRE (450 GW in 2030), at least prior to displacement in case a battery is chosen for NEW supply.

**Fig. 5 Chronological (half-hourly) 2030 Fuel Supply (before considering New Supply and consequent feedback loops)**

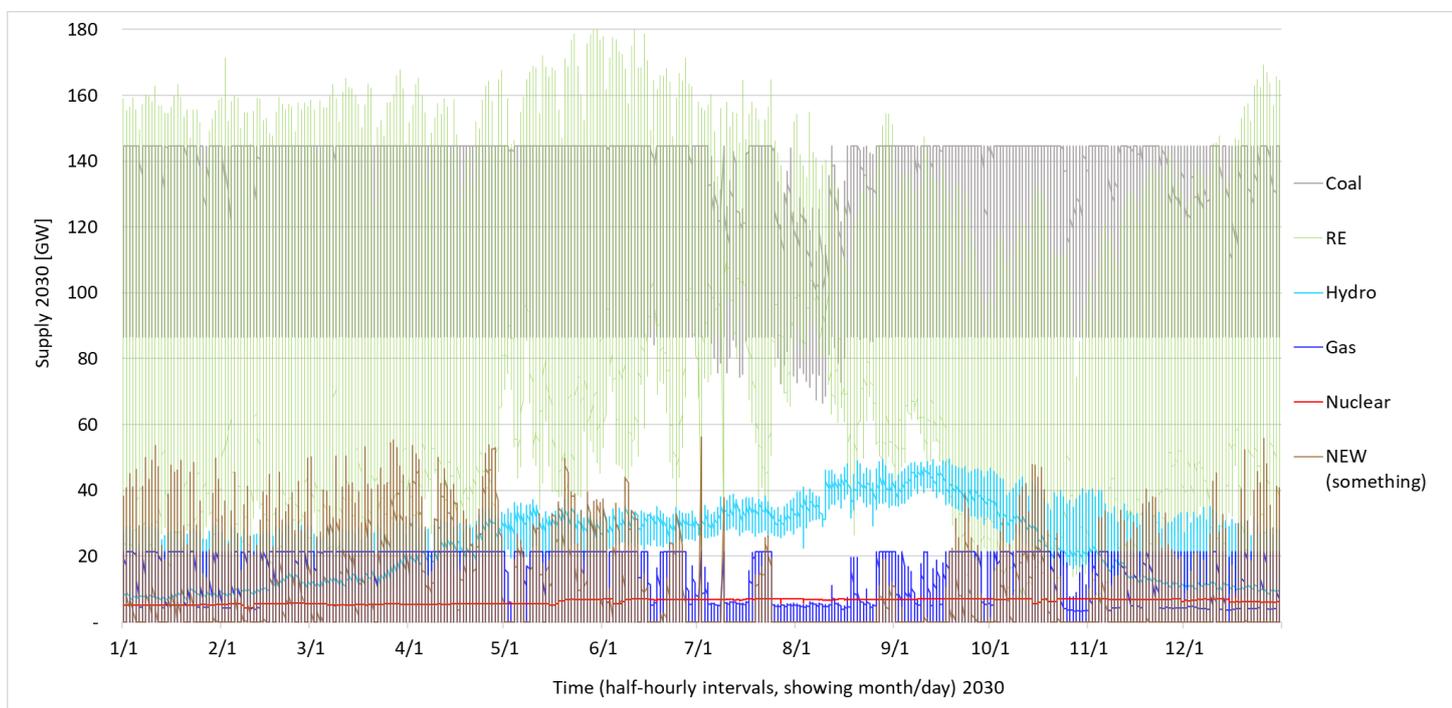

*\* The addition of NEW supply can change the volumes of other fuels, e.g., more coal added might displace expensive gas, or battery use beyond what is needed for unmet demand (NEW something) would displace fossil fuels. This effect is shown subsequently. Without data on alternate (non-power) requirements for water out of dams and expected future monsoon, given demand curves were scaled from 2019, hydro is also scaled from 2019. Future analysis will apply more variation in hydro duty cycles.*

*\*\* In practice, coal output wouldn't be as tightly within the bands shown (based on maximum possible and minimum required due to flex operations limits), especially because different plants would schedule maintenance differently.*

Renewable energy can grow measurably even as VRE without storage – 450 GW isn't too much. While high RE does have high curtailment, higher RE is still always a dominant strategy (lower costs, all else being equal, shown subsequently in Fig. 10). Curtailment levels heavily depend on the capability of coal plants to operate in a part load flexible duty cycle; Fig. 6 shows the impact of changing coal technical limits of flexible operations. There is much greater curtailment of RE due to coal limitations than due to simple surplus of RE compared to demand. More solar also leads to more curtailment.

**Fig. 6 RE curtailment in 2030 prior to feedback from NEW supply (a) before considering thermal flex (part load) constraints; (b) & (c) after accounting for thermal flex (part load) constraints with varying shares of solar**

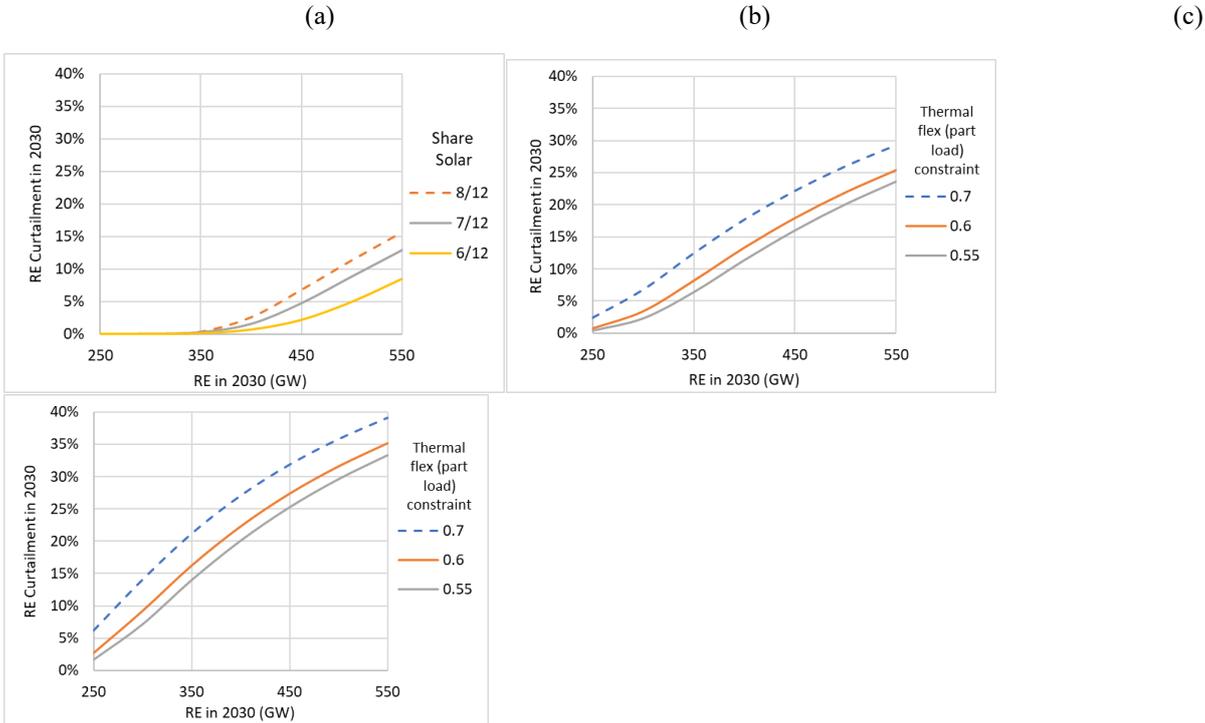

** (a) Assumes zero thermal flex constraint; (b) assumes solar:wind is 1:1; ratio (c) assumes solar:wind is 2:1 ratio

Seasonality can also be measured via curtailment of RE, which depends not only on variations in RE supply like for wind but also demand shifts).

While there is inherent uncertainty in RE output over time (as well as on demand load profiles), broad differences between solar and wind (diurnal variations versus seasonal) indicate immense economic value to higher shares of wind), despite higher per kilowatt hour costs. This is because at a portfolio level, higher wind reduces the need for "new capacity", even if there is greater curtailment during some seasons.

As Fig. 5 above showed, gas use rises, more so during peak periods, and coal use increases to the point of maxing out capabilities during periods of peak "net demand", again, before considering feedback from NEW supply. The supply of coal is limited by maintenance cycles and other downtimes, which we assume for the peak-demand period as an aggregate of 10% of capacity at any given time.[18]

This analysis emphasizes the need for technical design and planning to focus on instantaneous peak requirements instead of aggregate energy requirements. This also underscores the value of a smarter grid where balancing is not solely based on sufficient supply with a buffer, but also shifts in demand based on supply conditions. Saving electricity at the peak has not just avoided energy savings (Rs./kWh spread between expensive vs. cheaper), but also peak *capacity* savings (Rs./MW); LBL (2019) has data and a framework from the US. While the exact value proposition of avoiding the peak varies by system, it is a universal reality that the peak is always more expensive than the average cost of power, sometimes by several times or even an order of magnitude.[19]

---

[18] 85% coal power plant availability requirements per many power purchase agreements are aggregates, and one assumes planning/coordination leads to "planned maintenance" minimised during peak demand periods.
[19] This is based on first-principle techno-economic analysis, with a low PLF. Market prices at the peak can often be much higher.

## 5.3. New Capacity Required

Even very high growth of RE, combined with exhaustion of slack capacity from 2019, isn't sufficient to meet all the growing demand through 2030. This is because of not just a simple energy (kWh) calculation but (in)ability to meet instantaneous demand at each time. Growth of additional despatchable power (firm power) is required over time (Fig. 7).

**Fig. 7 Requirement for NEW supply (otherwise unmet demand)**

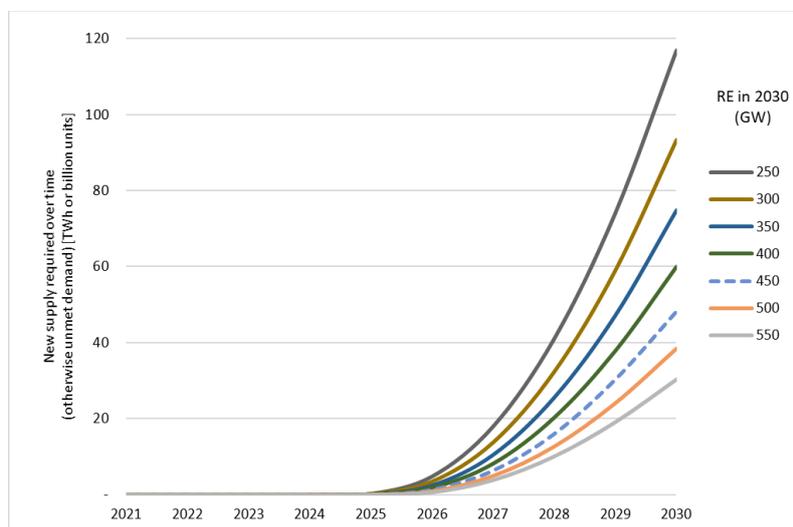

This new capacity will not be required at high capacity factors of utilization (or plant load factor, PLF, Fig. 8), and thus any new capacity regardless of type will end up being expensive.

**Fig. 8 CUF (PLF) of new supply on an energy basis (prior to feedback loops displacing existing fossil fuels)**

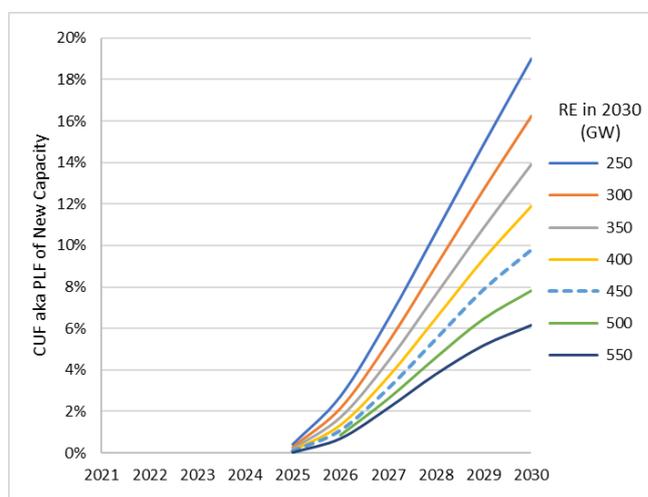

*\* This is simply the PLF of NEW capacity as required to meet unmet demand. NEW supply could operate more than this if the marginal costs are lower than the fuel costs of today, e.g., NEW coal displacing expensive (non-APM) gas, or batteries displacing both gas and coal.*

**Fig. 9 Load Duration Curve for NEW Supply required (half hourly stacked time periods) in 2030**

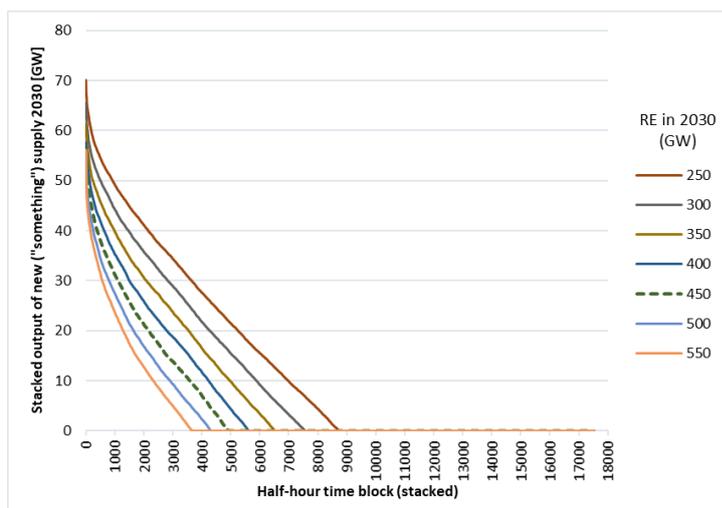

Fig. 9 shows the load duration curves (LDCs) of new supply, a graph that stacks output in a declining level across the time periods. This is after the growth of RE (as VRE) and also using up slack coal and gas capacity, and planned additions of hydro and nuclear power. This shows the required peak capacity for NEW demand as supplied is modest (for 450 GW base case and 5% demand growth, it is 49.7 GW).[20] The figure also shows this capacity is only required for part of the time, with zero "new" required for the majority of the time periods. The peak unmet load only varies between 70.1 and 56.1 GW across the wide range of 2030 RE. For the high RE scenario, the peak otherwise unmet (NEW supply required) size is similar to the capacity found for battery requirements by Abhyankar, et. al (2021).

## 5.4. Economics of the system – excluding sunk costs

Fig. 10 shows the NPV of the system based on future choices (level of RE and options for NEW supply) under base planning. Batteries are cost-effective under most scenarios primarily due to the feedback loop displacing fossil fuels (which become expensive by 2030). For some other fuel options for growth, higher RE is more expensive as curtailment grows without sufficiently reducing the sizing (capex) needed for NEW.

**Fig. 10 NPV of total system costs excluding sunk costs across choices of options for NEW supply and increasing planned VRE by 2030**

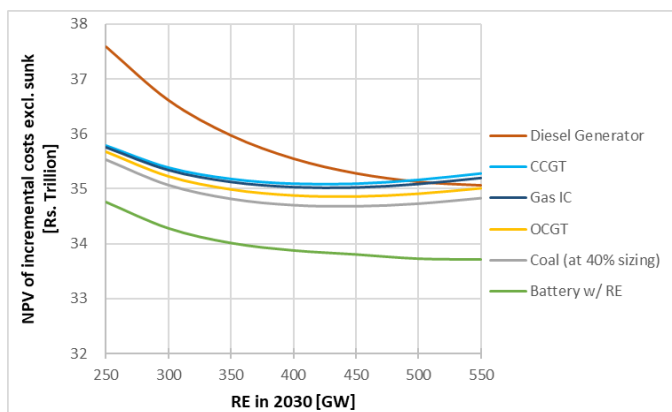

*\* Assumes 6% discount rate for NPV calculations, distinct from weighted average costs of capital of 8.5%*

*\*\* This is the incremental cost for RE expansion and operations, fuel for the existing capacity, and full costs for anything new added (capital, fuel, O&M). The RE with battery is separate dedicated solar. 40% sizing for coal*

---

[20] Any battery required would be slightly higher capacity due to depth of discharge limitations. Other fuel options would also be higher sized due to auxiliary consumption. They would also need to be sized to meet grid buffer capacity requirements.

*is compared to the full size to meet NEW supply (otherwise unmet demand), with the balance to be met by biodiesel.*

Most importantly, we find lowest costs for all scenarios and choices with higher VRE based on the lowest cost expansion technology (batteries with RE). Additionally, a higher share of wind is always cheaper regardless of VRE level in 2030 or fuel choice for NEW. While more wind increases the capex required (and even has a higher LCOE), not only is the output higher, but it also reduces the need for NEW supply because of its partial support for the evening peak. It also faces lower curtailment.

A key feature of this analysis is the parametric choice to undersize NEW capacity like a battery or coal compared to the peak NEW requirement, and have the residual met by, say, biodiesel (notionally zero carbon, but, more importantly, low capital costs offset by high operating costs). SI 4 shows the impact of choices on under-sizing NEW coal and batteries. Coal always benefits but batteries do not, because of the value of batteries to displace more expensive fossil fuels by 2030.[21] NEW Coal's optimal is close to 40% of peak NEW sizing – some of the savings are due to displacement of expensive (non-APM) natural gas in the future. Thus, up to 20 GW of NEW coal could be cost-effective, depending on how much RE is able to be built.

Under-sizing batteries from the worst time period requirement doesn't save as much money as linearly expected because it doesn't proportionally save RE requirements to charge the battery. There is often surplus (curtailed) RE which provides "free" charging of the battery. In addition, under-sizing the battery also lowers the volume of displaced fossil fuels (coal and gas) which help the use-case for batteries. However, depending on the shape of peak demand, there could still be a value for 5-20 GW of biodiesel *capacity*, even if rarely used, to displace peak NEW capacity deployments, and also as an insurance policy against unpredictable RE generation (or even a poor monsoon).

One methodological challenge is how to allocate fuel costs with NEW supply. The first few steps (modules) of iterative despatch analysis show the cost of all existing fossil fuels, leaving a blank for something NEW. Costs of NEW are then calculated, inclusive of feedback loops like displacing otherwise operational fossil fuels. However, the reductions in gas (by, say, coal or battery) are not shown as savings in gas fuel costs in the NPV calculations but apportioned to the NEW supply costs, which are now lower by this same amount. This helps understand costs segregated across existing fleet operating costs and planned RE vs. something NEW, even if the latter lowers the fuel used by existing plants.

**Fig. 11 NPV of costs of existing fleet plus planned RE (before adding NEW supply)**

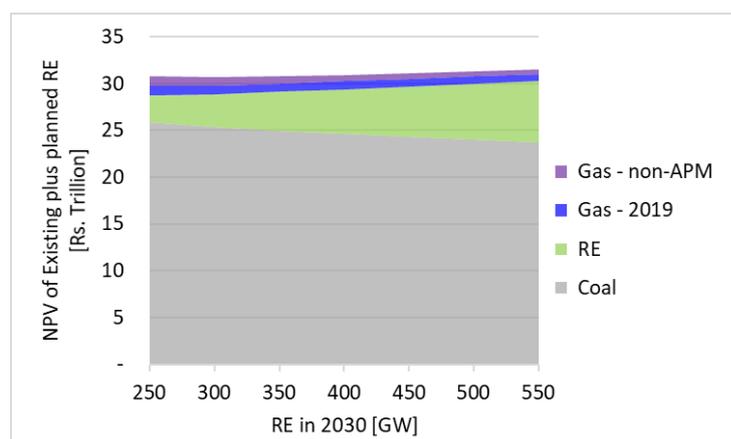

*\* These are not the total costs of the existing fleet before adding NEW supply. Some capital costs are treated as sunk (existing fleets) or exogenous, e.g., the costs of growing hydro or nuclear. Gas is split between cheaper*

---

[21] Use of a single price for fuel overstates the early cross-over for batteries over coal. If we recognize that some large fraction of the generation comes from pithead (and thus cheaper) power plants, even if the average future price of coal at the margin (fuel only) is higher than the cost of a battery, some fraction of the fleet remains cheaper. This will delay the ultimate crossover, and perhaps also change the estimate in the paper (based on averaged pricing of coal fleetwide) that higher battery capacity is always cheaper because it always displaces more expensive coal.

*(which ran in 2019) and more expensive supply for economic reasons shown subsequently. The cost allocations are before changes due to any NEW supply, which can lower fossil usage and thus costs of the existing fleet.*

If we consider the system costs for existing capacity and planned RE excluding sunk costs (Fig. 11) *before factoring in NEW supply or its potential feedback loops*, coal fuel costs dominate, even with high RE planned. However, this is the total expense for the year brought into present value, but the generation is vastly different across the fuel mix shown. Coal generates much more than RE in early years, which is the reason coal costs will inherently be higher. Fig. 12 shows the generation mix over time in a non-stacked and absolute manner. This also highlights the difference a battery makes if it is used to fill in the NEW. SI 6 shows the impact of different levels of VRE in 2030 affecting the generation mix.

Despite having sufficient coal capacity, there is still a need for NEW supply. This is because a simple kWh-based calculation of coal's capabilities vs. demand growth misses ToD aspects – "on average" there is sufficient coal, but not at all time periods.

Importantly, the total costs shown in Fig. 11 rise with higher RE. What higher RE does is reduce the costs for NEW something added to meet residual demand, captured in the *total* NPV of Fig. 10.

**Fig. 12 Generation mix over time with 450 GW RE (a) prior to adding NEW; (b) post adding a battery for NEW**

(a)                                                                                      (b)

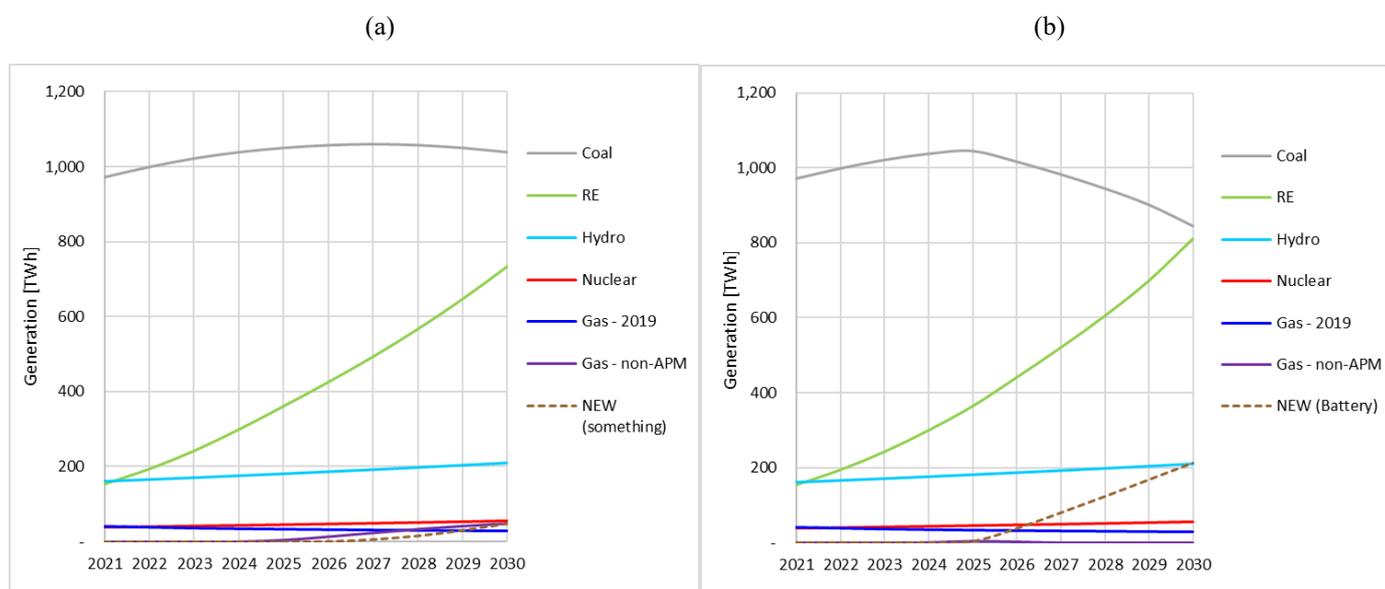

*\* The need for NEW capacity begins one year prior than generation needs due to the requirement of grid buffer capacity.*

*\*\* Without a battery, non-APM (expensive) gas rises more than cheaper gas because of time-of-day mismatches. If a battery is deployed, non-APM gas is used a for a few TWh in the middle of the decade, but then displaced entirely. This suggests a slight mid-decade acceleration of RE deployment may be beneficial. It also emphasises the importance of capacity, even if rarely deployed. Near zero PLF gas capacity would have nearly zero carbon emissions.*

*\*\*\* The RE share excludes the reduced curtailment due to the feedback loop with a battery, which is calculated as part of NEW instead. This also excludes any dedicated solar for charging the battery, which is seen by the grid through NEW (battery).*

If we consider levelized incremental costs, akin to LCOE, discounting both expenditure and electricity (since output varies over time), then the differential across the existing fleet with planned RE is much lower (Fig. 13). This is before feedback from NEW supply. The expenditure is based on amortized costs, i.e., if one needs a trillion rupees of investment in 2030 for a technology, the costs paid in that year are much lower, but then spread out over time, equivalent to a mortgage or EMI.

**Fig. 13 Levelised per unit costs for existing fleet and planned RE (excluding sunk, before NEW supply)**

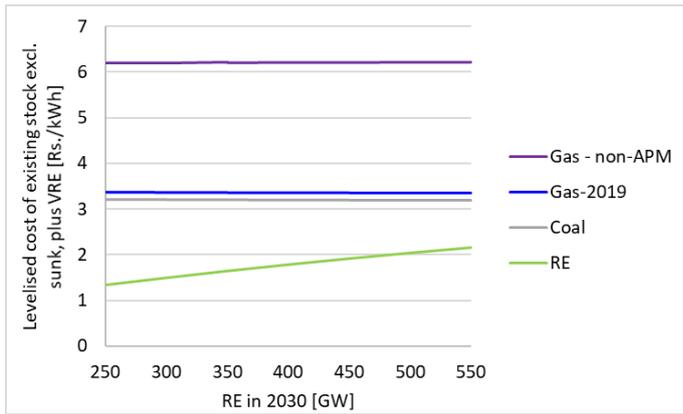

*\* Growth of hydro and nuclear are treated as exogenous and common across scenarios, hence not part of the analysis (akin to sunk costs). Capital costs of existing capacity are also not part of these calculations.*

*\*\* The RE per unit costs rise with higher RE levels because more RE by 2030 would require more capacity up front, before costs decline (and discounting further reduces its present value).*

The volume figures do not show the RE required to charge the battery, which can be very substantial (SI Fig. 21). For 450 GW of RE with 2:1 solar:wind, we need 93 GW of solar for the battery. Such capacity could be shown separately from VRE (which is modelled as exogenous targets, e.g., 450 GW), but such dedicated solar's *energy* doesn't show up directly, rather it charges the battery which is used to supply NEW output.

Similar levelised costs for NEW supply show a stark difference amongst options (Fig. 14). Most options are very expensive, in part due to the very low PLF of the new capacity,[22] but batteries display a low cost. This is because the costs here are not only discounted, but also reflect a cost reduction allocated to batteries from displacing fossil fuels. This is because the analysis focuses the incremental costs of different NEW options, and all the benefits of lowered expensive fossil fuels are allocated to the NEW (battery) supply.

**Fig. 14 Levelised per unit costs for NEW supply**

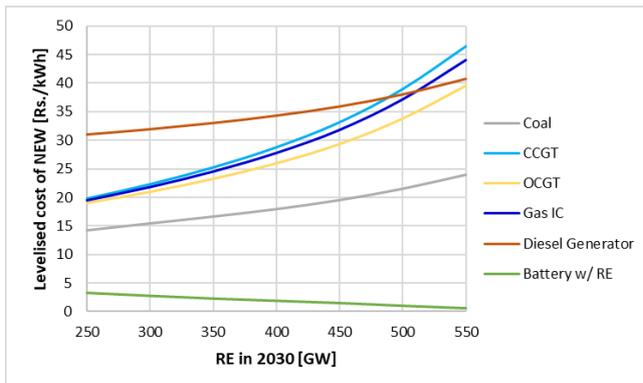

*\* High costs are primarily due to very low PLFs for NEW.*

*\*\* These costs reflect offsets (feedback) displacing fossil fuels. Without feedback, the standalone costs are higher for coal and batteries. Batteries also assume harnessing spare RE for free that is otherwise curtailed.*

---

[22] The present surplus of capacity is one reason for the low CUF of NEW. Fig. 8 shows rising PLFs over time, so the costs will fall per unit for any capacity. However, such growth will be in the future, where fossil fuel costs rise as well. These factors minimise the limitations of not modelling beyond 2030.

## 5.5. Batteries are cost effective under cost trend projections

An important and cost-effective NEW supply option in most scenarios is the use of storage technologies such as batteries; this assumes pricing declines as projected combined with a rise in fossil fuel prices. The model does not distinguish between use cases or duty cycles for the battery, such as whether it is designed to avoid curtailment of VRE or used to meet residual unmet (peak) demand. By applying a parametric analysis, we can quantify the lowest cost portfolio where both uses are maximized.

Recent studies have aggressive costs for batteries in 2030, e.g., Rs. 3.7/kWh in Abhyankar *et al.* (2021). These are levelized costs of energy (LCOE). Without feedback, we find the effective cost is demonstrably higher because of the duty cycle, expanded below. Given the fact that batteries have virtually no operating costs, underutilized capital is expensive. Batteries need a source of power to charge, which we assume to be renewable energy to meet decarbonization objectives. There is usually sufficient "surplus" or curtailed RE on a daily basis to power the battery but this varies by level of VRE deployed by 2030. SI Fig. 21 shows additional RE required to meet NEW demand beyond using spare (curtailed VRE). Seasonality effects dominate, and the period when there is the highest surplus RE (high wind) is when the battery isn't required, at least to meet an unmet residual peak.

The sizing of the battery and requisite solar are back-calculated based on the intersection of energy and capacity requirements at every timeslot. We decouple capacity and energy such that it would be possible to have one kilowatt of output for a wide range of hours (not just 4 hours of storage). Fig. 15 shows the required size of batteries to meet all the unmet residual load on a kWh basis. Any charging of batteries beyond using curtailed RE is from dedicated solar power, which gives relatively predictable daily charging.

These volumes of battery are higher than found by Abhyankar *et al.* (2021), who ran an optimisation that found higher coal capacity than our base for 2030. These findings are in sync because undersizing the battery creates relatively low levels unmet demand (see SI 4 for details on undersizing the battery).

**Fig. 15 Battery capacity (kWh) required over time to meet *entire* residual (otherwise unmet) demand**

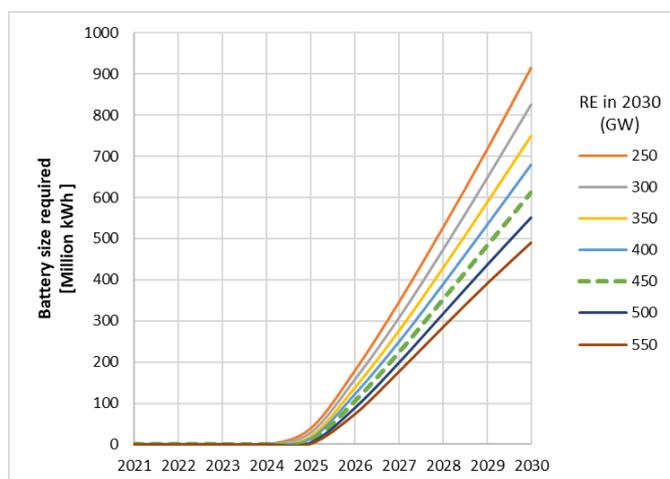

\* Assumes 90% battery efficiency, and 5% minimum depth of discharge buffer. The capacity requirements can be estimated from Fig. 9 (after adding in depth of discharge and buffer requirements). This assumes no other capacity growth beyond planned RE, and modest growth of hydro and nuclear. Small growth of coal (e.g., within presently under-construction plants) would lower the battery requirements. This figure also doesn't undersize the battery to optimise costs by reducing size compared to worst-case-demand battery requirements, shown separately (SI 4).

By separating battery energy from inverter capacity, we optimize based on separate requirements. We find the peak requirement is well more than 4 hours battery. If we consider 2019 shape curves and scale these, 35% of demand fell within the RE shape curve. This helps estimate the size of the battery required, at least before changes in solar:wind ratios. With solar producing, say, 25% output, or 6 hours per day, this means 4 hours is available to charge the battery if about 1/3 matches demand. Because of daily variations, increased dedicated solar for charging a battery has limited value as much of it becomes curtailed (when the battery is full).

The battery with inverter capacity (MW) required can be read off the NEW LDC in Fig. 9. For 450 GW of RE, the "new something" net requirement peaks at 56.13 GW of unmet demand in 2030, assuming a 5.25% demand growth. The NEW something in that figure is net, so increases based on auxiliary consumption or depth of discharge limitations would apply for respective fuels. Adding in grid buffer requirements as well, we find that for 2030, the inverter size required 68.7 GW (with a 2:1 solar:wind ratio). This capacity along with the commensurate kWh of storage aren't required every day to meet the residual peak, and thus other days the battery may be underutilized.

Batteries can have several duty cycles of usage which strongly influence their system level costs. This is an important distinction of this analysis compared to literature which either assumes specific pairing of energy and capacity (like a 4-hour battery) or treats battery like a fuel at a specified levelized cost of energy (LCOE), such as 4 or 8 Rs./kWh. Analysis based on blending gives a low cost for RE plus batteries. If it assumes Rs. 2.5/kWh for solar, and 8 Rs./kWh for storage (LCOE), but then assumes only 25% battery requirement, such calculations lead to Rs. 2.5 + 25% x 8 = 2.5 + 2 = Rs. 4.5/kWh. However, this is incorrect because one doesn't need a quarter sized battery, but rather one that is larger for occasional needs and ends up as a quarter only on an energy basis.

The duty cycle from batteries leads to several possible scenarios of utilization. If the battery is sized exactly to meet the worst-case conditions of the year, then on other days of lower residual demand it isn't required in full at least to meet NEW supply. In such a case, given it has already been deployed, the battery would be utilized and displace other supply with higher marginal costs, and thus save on fuel costs. However, in such a case, its value is not equal to the total cost but only the marginal fuel cost avoided. For the kWh where batteries displace coal, the marginal value of in 2030 could be close to 4 rupees per kilowatt hour for national coal fuel prices. Batteries become cost-effective for such use by first displacing more expensive natural gas, and also due to a secondary feedback loop that lowers the daily high coal use, which reduces curtailment.[23]

Given such a sizing of a battery is expensive, we parametrically assume a range of battery sizing compared to the instantaneous maximum peak requirement. We find that high (rather, full) battery sizing is cost-effective due to high fossil fuel costs projected, but optimal mixes depend on the level of VRE planned and wind:solar mix, and are non-linear.

Fig. 16 shows the state of charge of for fleet batteries in 2030 based on, as an illustrative example, an 80% sizing compared to the unmet demand (full sizing) and assuming it were entirely charged daily. Charging could be from a combination of surplus RE that was curtailed and/or additional solar RE for charging the battery, which has a cost. The utilisation for this figure is only for NEW, and before considering feedback loops where a battery would be used to displace fossil fuels.

**Fig. 16 State of battery charge over time with 450 GW RE (as VRE) if only used for otherwise unmet demand ("something new" required) with 20% kWh under-sizing compared to full unmet demand**

---

[23] In the very long run, there will be limited feedback displacement value, and battery + RE economics will need to be standalone competitive. Batteries are lowest cost in part because *all* NEW solutions are expensive, and once built they should be maximized (with no fuel costs).

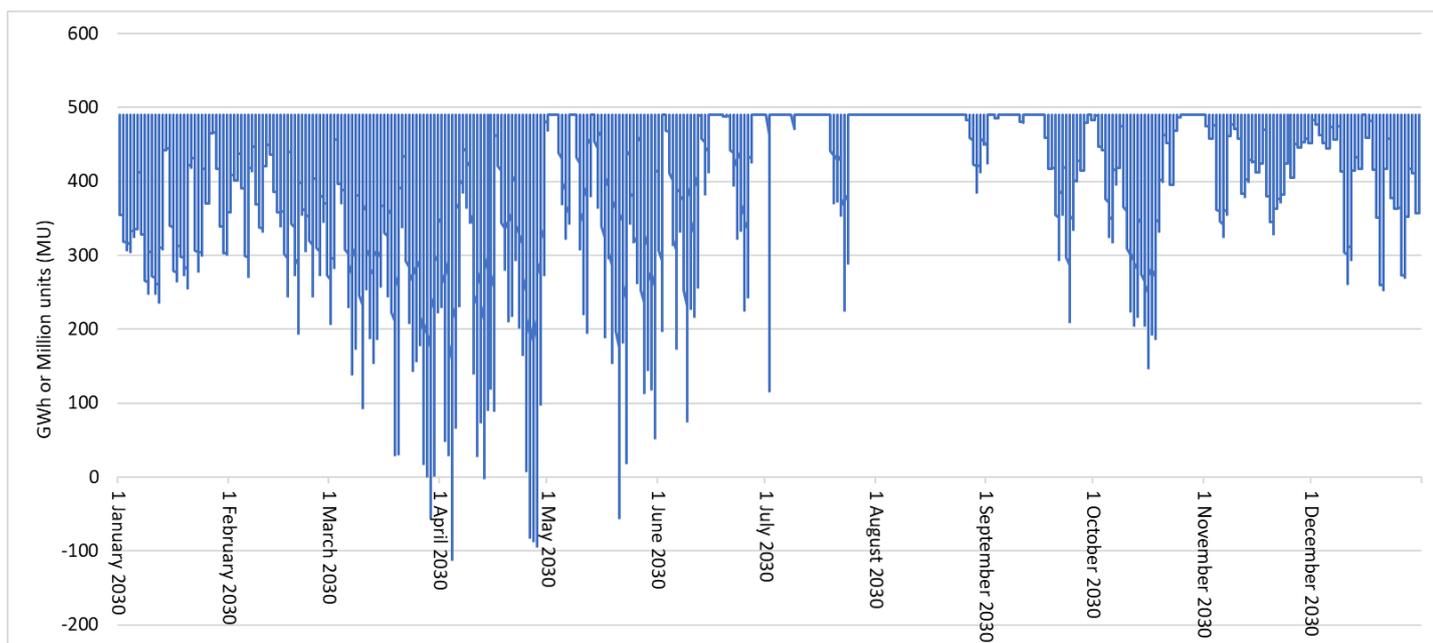

*\* Shown are half-hourly data points across 2030. Negative values mean insufficient battery, with residual (secondary) unmet demand to be met by other means, such as biodiesel. Battery usage would be limited not just to positive states of charge, but also the depth-of-discharge limit (95% in the base).*

*\*\* If one sized the battery to full, it would just reach its minimum (depth of discharge limit) at the worst-case point in the year.*

*\*\*\* This duty cycle is before using surplus battery charge to displace fossil fuels, calculated separately.*

Most days the battery isn't required in full, and there is a large period in the monsoons where it isn't required to meet unmet demand at all. Only occasionally does an undersized battery fall short. Of course, the battery would be used other days as well, to displace fossil fuels, but its value would only be avoided fuel costs. The economics shown in Fig. 10 are inclusive of chosen supplemental solar, and using the "spare" battery to displace gas and coal on a daily basis.[24]

If one undersizes NEW capacity to save battery (or even coal) capital costs for infrequent peak NEW demand capacity, this leads to a very modest or low secondary unmet energy requirement (SI Fig. 19), but a higher savings in capital costs. Halving the battery only leaves 8% residual demand unmet in 2030, for the base case. This could be met by, say, biodiesel (assuming one wanted a ~zero carbon solution with low capital costs). However, the only reason this doesn't lead to lower total costs for batteries is because high future gas non-APM costs mean a full-sized battery is still cost-effective for the reasons given before, but such options must be revisited with a sharper analysis of peakiness of demand and also a better understanding of fossil fuel prices in the future.

Once the average cost of batteries becomes lower than the marginal cost of coal, this would mean a decline in existing coal utilisation as optimal.

How much battery is spare is calculated conservatively on a daily basis by examining the minimum state above the depth of discharge limit and also comparing that against the surplus in the day where the battery becomes full but there was still some charging capability through a combination of usable curtailed RE (from the planned VRE) or dedicated solar planned for the battery.

---

[24] The displacement of fossil fuels by spare battery capacity is calculated not on a chronological day basis but on a 24-hour cycle post *full* solar charging, i.e., today's solar charging can operate overnight and through the subsequent morning. This also assumes sufficient clean energy to recharge the battery in full daily (through a combination of otherwise curtailed RE or new solar added explicitly for the battery.

## 5.6. High RE is optimal

While all studies are based on assumptions, and have different methodologies, our findings reinforce other findings that high RE is cost-effective for meeting India's upcoming electricity requirements.

The lowest cost solution gives the following insights and trends:

1) More RE even as VRE is cost-effective; higher wind also helps lower portfolio costs for all fuel types for NEW supply.
2) Meeting unmet demand ("NEW") is expensive regardless of the choice of technology or fuel. This emphasizes the need to manage demand and reduce peak net demand. High VRE gives a positive feedback loop near the end of the decade if batteries are deployed because curtailed RE can power the battery for free.
3) The existing fossil fleet is cost effective, even at high gas prices, when used as a peaker.

While changes in load profiles and supply shapes will impact the specific optimal mix, the trends and general directions are consistent.

High RE leads to "peak coal" (supply plateau) by approximately 2027-28 even without a battery (which accelerates the timeline) because VRE displaces coal for much of the day. In fact, generation rises in the short run, before plateauing and declining under high RE scenarios. Fig. 12 has more details, including with a battery and SI 6 focuses on coal. This is assuming high RE CUFs for new builds, else peak coal is delayed.

In this analysis, we do not have a module for growth of coal except where required to meet otherwise unmet demand (NEW something required). However, by varying the retirement rates for existing stock, we find it much cheaper to maintain as much of the stock as possible even if it has a reduced PLF. Policies aimed towards air pollution could limit their capacity and/or output, but peak capacity reduction is worth much more than average (energy) reduction. If under-construction coal power plants are completed at much lower incremental cost than greenfield, then this could be worthwhile, estimated to a level of a few 10s of GW.

## 6. Limitations and Further Studies

Because this paper focuses on understanding uncertainty and implications of choices, it is, by design, not a full load flow analysis with spatial resolution. It aggregates at an all-India level, which in some ways could be considered the best-case scenario, ignoring transmission costs and limitations. It is also not a unit commitment analysis, instead focusing on balancing and national level economic despatch. While the analysis is, like most literature today, deterministic, the explicit treatment of a wide range of uncertainty gives robustness to the results, as well as insights for further studies. Because the objective is to understand decision choices and uncertainty, the economic analysis isn't for full system costs, but only non-sunk future costs based on choices such as RE capacity in 2030, ratio of solar:wind, choice of supply for meeting NEW demand, under-sizing the battery, etc.

Other specific limitations include:

1. The use of a single year, 2019, for supply and demand profiles. These will obviously be evolving as well as stochastic over the years. Unfortunately, such data are not easily available at high time resolution.
2. Future RE is assumed to have a high CUF (PLF), which is possible and already the norm per discussions with developers. However, if such higher PLFs do not continue for the entire growth, then RE output would be lower, and coal demand would be correspondingly higher. This is one reason to use this analysis for trends and illustrative use, and not predictive modelling.
3. We apply linear scaling of supply and demand curves over time based on parametrically varying growth rates. As and when proper time series data are available, this can be enhanced. In addition, technology and policy shifts can enable both load profile changes and supply changes. The government has already embarked on a massive program of solarization of irrigation pumpsets, and the rise of smart grids could dramatically increase flexibility in balancing the grid.
4. The planning horizon is limited to 2030, even though any new capacity investments would have a remaining lifespan well beyond 2030. However, high discount rates in the Indian scenario reduce the impact of extending the horizon. Rapidly evolving price points for clean technologies make it difficult to project well beyond 2030.

5.  Growth of RE is assumed to be rising over time, instead of equal (linear) per year to meet chosen 2030 values for RE. There might be a likely push up front for RE, but it is unlikely to make it front-heavy or even linear. However, the RE growth trajectory impacts fossil fuel demand (and even how soon one needs something NEW).

6.  While part of the study emphasizes the difference between wind and solar, these were estimations splitting aggregate RE because segregated data is unavailable. We relied on NREL's PVWatts tool to consider a prototypical (mid-India) solar profile only to partition the total RE across wind and solar (other RE is small); we assume limited impact of "other RE" whose targets have limited growth through 2030. We simultaneously projected aggressive PLFs for solar and wind for future builds, at levels much higher than the average today. These were based on discussions with leading developers. The actual output could not only vary stochastically but be heavily dependent on location, especially for wind.

7.  We did not apply ramping constraints explicitly but rather checked post facto whether ramping rates were in compliance with aggregate norms. There are periods of concern, and it remains to be seen if practical ramp rates match theoretical potential, varying by age of plant and technical design, as well as load distribution. For the base case 2030 with 450 GW RE, only 5 half-hourly time slots need a coal ramp rate above 1% per minute (and below 2%), and only a few percent of time slots are between 0.5% – 1% ramp rates, calculated on the basis of nominal capacity in operation that day.[25]

8.  While we did consider part battery and also under-sized coal solutions for meeting the peak blended with biodiesel, we did not consider fully blended portfolios of all new capacity or multi-fuel optimization.

9.  Due to computational limitations, and large n-dimensional space used under parametric analysis, we limited ourselves to 30 minutes resolution.[26] Even 5-minute ramping is very different for the grid, before considering transient power flows and 3-5 cycle long events (tens of milliseconds). Computational limitations also make analysing blended portfolio options for NEW supply more difficult. While in theory there are infinite choices when one adds combinations, techniques to handle this include use of dominant sets, bounding, and clustering.

10. The national aggregate picture masks heterogeneity and differences at a regional, state, and plant level. The true picture would inherently be more expensive than an optimal national picture that ignores such granular constraints or transmission. However, there is no evidence that the trends or policy implications would be substantially different.

11. The analysis doesn't focus on grid stability issues, which will be important as we will have heterogeneity across the nation in terms of plant despatch. For example, a 10% reduction in coal supply would not necessarily mean all plants reduce by 10%, and hence there would be load-flow implications worth examining further. The use of Phasor Measurement Units (PMUs) at the transmission level will help measure real-time state estimation. These would increase our ability to safely stretch the grid to its limits, which improves resource utilisation and thus saves money. Otherwise, the traditional solution has always been over-engineering.  Such grid stability analysis should be combined with plant level or at least state-level analysis.

12. In addition to technical uncertainties (including stochasticity), there are vast uncertainties in prices over time, especially for rapidly evolving technologies including RE and storage, but even for fossil fuels. These don't just affect total costs but cross-over points between technology choices. These are compounded by policy uncertainties which could both shift price points as well as create non-technical constraints such as contractual, e.g., an obligation to use a certain volume of a particular fuel or supply, or a carbon price. We also do not factor in issues of construction timelines for optimal planning.

---

[25] While ramping is likely not an issue in aggregate, in practice selected plants may face constraints given the average output will be unequally distributed across plants and locations. Some plants that were very low or high in output may need disproportional ramping, and if this approaches their limit, the alternative would be pre-emptively distributing the ramping needs across more plants, which would have some (but likely small) economic implications.

[26] Half hour resolution is 17,520 data points, and if we only consider 5 parallel variables with 3 possible values ($3^5$), that alone is 243 dimensions in parallel. Each of these time points has iterative cycles of economic despatch, leading to analysis across millions of data points.

This model should hopefully be a steppingstone to deeper studies with greater time series data as well as creating stochastic models. An ongoing study by Aniruddha Mohan *et al.* (under preparation for publication) based on 20 years of MERRA-2 satellite data for wind and solar output in the Indian context shows a little under 10% increase in impact on portfolio planning economics due to capturing "outlier" weather patterns; Staffell and Pfenninger (2016) give a reanalysis framework.

# 7. Conclusion and Discussion

## 7.1. Summary of findings and implications

1) Higher RE as variable RE (VRE) is the optimal strategy for India's grid. 450 GW RE by 2030 (even without storage) is ambitious, but cost-effective, even with high curtailment;
2) Adding high CUF wind is important at a systems level, even if the LCOE is higher than for solar;
3) Even with very high RE and minimal coal fleet retirement, modest demand growth for electricity of 5.25% means there is insufficient supply at many time periods, and so "something new" is required;
4) India's grid has benefited at an operational level from surplus coal capacity, which will exhaust in a few years. 2030 simulations show maxing out of existing capacity, and a rise in gas output. Thus, it has not yet faced high marginal costs for peak power, which will be what most of NEW supply will be for. A corollary finding is that India needs to ensure adequate fossil fuel supply;
5) NEW supply will inherently be relatively expensive because it will operate at low PLFs if its duty cycle is to meet otherwise unmet demand, or because its value is limited in some time periods to the savings based on displacing fossil fuels. Stated another way, India will need to embrace the concept of peakers in the future, which have low utilization and higher per unit (kWh) costs.
6) In case a battery is used (with RE), it is the cheapest of NEW supply options not because it is cheaper than coal on a standalone (LCOE) basis but because it can displace expensive fossil fuels, and so offers value beyond simply meeting NEW supply (otherwise unmet demand). Batteries also benefit from the ability to use surplus (curtailed) RE, modelled at marginal costs, i.e., free RE; This is one reason we find high energy (TWh) coming from batteries by 2030 – much of it is because the capacity displaces fossil fuels in some time periods when otherwise not required.
7) A disproportional share of incremental costs is for selected (peak) time periods, especially when measured by "net demand" (firm demand after removing RE supply). Time of day pricing will be important for both shifting consumption to off-peak periods and also signalling for the correct investments in capacity. This includes the need for a smarter and resilient grid under uncertainty and volatility, which will only worsen as RE's share increases. The use of ToD pricing for wholesale procurement is likely to impact VRE more than traditional (despatchable) supply, and based on the coincidence of solar across most of India (especially compared to wind), ToD will ultimately reduce the value of solar mid-day;
8) India will need to work backwards based on construction schedules to see by when it wants to procure for NEW supply. By this time, there would be greater clarity on fossil fuel prices and technology prices, especially for RE and batteries. India should be cautious in premature retirement of existing capacity, even if this is rarely used, as it provides disproportional value at the margin.[27]

## 7.2. Discussion

The near-term economically optimal strategy (as well as current government policy) is to focus on adding high volumes of VRE (solar and wind). It's worth reiterating this is planned RE as VRE, without a battery, and if a battery is chosen to meet NEW supply, a substantial additional and dedicated RE deployment (likely as solar) is required to charge the battery beyond what curtailment can offer (SI Fig. 21).

---

[27] Singh and Tongia (2021) emphasise the value of an integrated framework for considering coal power plants and their retirement versus upgrades, both for pollution control technologies and flexible operations. There can be synergy to doing both upgrades simultaneously.

For achieving high planned RE, lowering the cost of electricity from these sources is the obvious requirement, whether through technology improvements or cheaper financing, but there are additional technical design issues at a system level that will need to be addressed over time including optimal location, especially in terms of synergizing with transmission or edge-based generation, as well as a strong need to consider time of day aspects. Even without explicit time of day pricing, procurement for electricity is likely to increasingly reflect times of day including through bids asking for higher PLFs, as well as bids with explicit peak period power requirements.[28]

This analysis finds strong value to increasing wind, but the LCOE of wind is higher than for solar. This means traditional bids are insufficient to reflect portfolio or system-level costs (even before considering transmission). Very high wind installations would also face limitations of appropriate land, ones with high wind speeds. This suggests the need for a strong wind re-powering program that might require buying out historical wind asset owners, whose older turbines sit on prime land. This analysis shows funding might be available from portfolio level cost savings by increasing wind's share in the future growth of VRE.

One of the key techno-economic and policy issues for rising VRE is how to handle curtailment. The increase in average costs is modest and RE even with curtailment is still cost-effective compared to alternatives for growth. However, would the economic implications be spread equally amongst all RE generators? All power generators? More importantly, in this analysis we assume curtailed RE is available at marginal cost for charging the battery, i.e., for free. It's not clear what policy or market instrument will make this happen. India lacks a market mechanism that signals marginal costs for most transactions. Power exchanges, excluding traders, only supplied 4.1% of power on market principles in FY19-20 (CERC 2020). They have largely had pricing determined by fossil fuel costs, especially coal.

Even with very high RE addition, rising demand will necessitate the need for additional capacity for firm power, which could be met through a battery or traditional sources or diesel/biodiesel; this is complementary to the use of demand response or Virtual Power Plants. Utilities in India have already aggressively promoted Demand Side Management (DSM) but such programs have mostly lacked a temporal signal – they focus on energy efficiency.[29]

Any such additional capacity would be better served by improved pricing signals that goes beyond average price under an LCOE metric. For consumers, Time of Day pricing can enable time shifting but these require not only appropriate tariffs by the Regulators but also appropriate digital metering infrastructure.

Pricing will also have to reflect system level costs including transmission, efficiency penalties on other generators when they back down (and also higher pollution, e.g., gas has much higher NOx under part load ramping (Apt and Katzenstein 2011)), and standby capacity requirements which will not be utilized often but still be required. India needs to enhance its basket of ancillary services in the power grid, beyond today's frequency support ancillary services.

As the analysis shows, all NEW supply will be relatively expensive due to low volumes of otherwise unmet supply. An exception is batteries because these can displace more expensive fossil fuels. However, the individual entities who provide peaking power as well as the utilities or consumers paying for the same will not be equally distributed. Hence, India will need to prepare for future with greater differentiation of costs based on engineering economic rationales instead of social welfare policy. *The key insight from this analysis isn't the comparison based on "are batteries cheaper than coal?" – it's showing they are the cheapest option for NEW supply.* This doesn't mean we should be shutting down existing coal capacity, because existing coal capacity is likely to be cheaper than addition of batteries, at least for a range of time periods when there is insufficient "free" (curtailed) RE to power the battery.

---

[28] Recent moves for RE are hybrid wind and solar or geared towards "firm power" through "Round The Clock" (RTC) bids. However, most RTC bidding to date hasn't been with storage, but, rather, oversizing the RE (Gambhir *et al.* 2020). This is the reason the bid prices are attractive but true long-duration storage-based solutions are still expensive, more so for states with surplus coal capacity and hence only the marginal costs of coal to be paid for incremental supply.

[29] The greatest success story for DSM has been with the use of efficient light bulbs, first CFLs, and subsequently LEDs. The synergy here was these disproportionately avoid evening demand, which helps the grid the most.

Simplified models that focus on LCOE ignore the time-varying value of an output. As this analysis shows, supply isn't meant to produce a flat or steady output, but one that varies over time based on both other supply and demand. LCOE is only a marker of value of a fuel choice, but a more expensive option based on LCOE may be cheaper at a systems level based on grid conditions, not to mention transmission, other constraints like ramping, etc.

While time of day pricing is a well-known need for the Indian system, it will ultimately require upgrades to not just generation bidding documents (procurement) but operational procedures at the load despatch level. Wholesale ToD prices will also need conversion to consumer (retail) ToD pricing, which is today limited to bulk users in India, and sometimes voluntary for them. Deployment of suitable metering technologies will be important for such consumer signalling.

While ToD pricing reflects the need for peak pricing, which will carry a premium, a corollary is the need for paying for resiliency, including a grid buffer. Today, surplus coal power capacity provides slack, but in the future, idle or rarely used capacity will need specialized pricing, which might need to be spread across the system (i.e., socialized). Much of the costs for NEW supply are for outlier time periods. Technologies to avoid such peaks can help dramatically, such as demand response.

Peak load disproportionately impacts system costs and unfortunately this places limitations on presently available tools for shifting evening peak demand. Assuming this continues to be driven by air conditioners for residential use (and partially commercial use), rising wealth, availability of solutions, and even climate change are likely to increase the magnitude, frequency, and duration of the spikes in evening peak demand. This pricing analysis for NEW supply can net-back the value of peak pricing and load shifting by offering an upper bound for the value of such options compared to simply adding NEW supply.

Half of India's coal fleet is under a decade old. Given the cost-effective energy available from existing fossil fuel capacity, India will have to improve grid operations and planning on a portfolio basis while simultaneously improving the technology and performance of its coal power plants. Not only must they be cleaned up (e.g., to comply with more stringent pollution norms by the Ministry of Environment, Forests, and Climate Change announced in 2015 (MoEFCC 2015) but not yet implemented by most plants), coal plants must increase their technical flexibility to operate at part loads and under high ramping conditions. While the specifications are for 55%, newer NTPC plants have proven they can operate at 40%. Germany's coal fleet showed actual ramping capabilities varying by vintage between 0.8%–4.5% per minute (Deloitte, 2019). The costs and benefits for such technical upgrades as required need to be evaluated holistically. While natural gas has a higher ramp capability in percentage terms (estimated at 5%/min as a minimum), the low base of gas capacity operating at a given time means the aggregate ramping from coal will continue to be much higher in India (Tongia 2021).

One spinoff result from this analysis has implications for green hydrogen – hydrogen produced from clean electricity through electrolysis of water. A significant component for green hydrogen is for electricity, and conventional wisdom has posited a synergy between a high RE world and green hydrogen. As this analysis shows, even a very high level of RE deployment will have limited surplus energy that is curtailed and thus free as an input for hydrogen production. Much of this surplus would also be needed by batteries if these provide NEW demand, and hence green hydrogen would likely require dedicated RE.

India will need vastly more nuanced and granular studies for optimal grid planning that not only increase the time resolution but also spatial resolution, eventually down to generator level along with inter- and intra-state transmission.

The biggest need will be for both modelling as well as technical understanding of variability and stochasticity of the electricity system. This is not just for variable RE, but also for demand, and even annual variations in hydro output. The latter is shifting further due to climate change as well as land changes that affect drainage and watersheds. Enhanced modelling requires extensive data, which has been difficult for researchers to access. Most studies have relied on satellite models or secondary data, which is a good starting point but insufficient.

Handling uncertainty must also factor in risk tolerance and risk management capabilities. How would grid buffer requirements need to change when the grid is heavily VRE? Grid operators, regulators, and utilities will need to converge on how to signal, incentivize, and sign up for additional capacity, respectively, for grid security and resilience means. Such planning has been missing in pricing based entirely on energy (Rs./kWh).

A key means to handling uncertainty is periodic revision in national plans factoring in the time constants of construction of different technologies and infrastructure including transmission and fuel supply. By the mid-2020s, we would also have a much better handle on clean technology prices as well as demand trends.

High RE as VRE remains the optimal strategy regardless of fuel and clean technology prices. However, not only is there uncertainty not modelled in this analysis, if storage costs don't fall as envisaged, and fossil prices also don't rise as fast, there are scenarios where the optimal solution for NEW deployment is reduced size battery storage complemented by biodiesel.

Fossil fuel prices are especially important for choosing new technologies because the latter can displace existing generation, especially with batteries. We assume modest rise in fossil prices (and higher rise for coal) (SI Fig. 18), but natural gas outside Administered Pricing Mechanism (e.g., via LNG imports) starts expensive and is thus much more expensive by 2030. In the future, if coal is the choice of NEW supply, it benefits from displacing such gas, one reason that even at low PLFs for NEW coal, the costs are lower than for natural gas, something that wouldn't happen otherwise.

Not only must modelling and analysis improve, policy-makers must be engaged to understand what are the decisions choices (and by when) for optimal fleet deployment and operations. There also need to be complementary analyses related to pollution, contracts, market design, and innovation, and their interplay with optimal grid capacity expansion and operations.

# 8.  References

Abhyankar N, Deorah S, Phadke A (2021) Least-Cost Pathway for India's Power System Investments through 2030. Lawrence Berkeley National Laboratory. https://eta-publications.lbl.gov/sites/default/files/fri_india_report_v28_wcover.pdf.

Apt J, Katzenstein, W (2011) Thermal Plant Emissions Due to Intermittent Renewable Power Integration. U.S. Department of Energy. https://www.osti.gov/biblio/1556903.

BNEF (2021) BloombergNEF's annual battery price survey finds prices fell 6% from 2020 to 2021. https://about.bnef.com/blog/battery-pack-prices-fall-to-an-average-of-132-kwh-but-rising-commodity-prices-start-to-bite/, Last accessed December 4, 2021.

Boyd E (2016) Overview of Power Sector Modelling. US Department of Energy – Office of Energy Policy and System Analysis. https://www.energy.gov/sites/prod/files/2016/02/f29/EPSA_Power_Sector_Modelling_020416.pdf. Last accessed Dec 5, 2021.

CEA (2019a) Flexible Operation of Thermal Power Plant for Integration of Renewable Generation. January 2019.

CEA (2019b) Monthly Executive Summary Reports. (multiple months) https://cea.nic.in/executive-summary-report/?lang=en. Last Accessed August 19, 2021.

CEA (2019c) Long Term Electricity Demand Forecasting. Central Electricity Authority, New Delhi.

CEA (2020) Report on Optimal Generation Capacity Mix for 2029-30. Central Electricity Authority, New Delhi. https://cea.nic.in/old/reports/others/planning/irp/Optimal_mix_report_2029-30_FINAL.pdf.

CERC (2015) Report of the Committee on Spinning Reserve. https://cercind.gov.in/2015/orders/Annexure-%20SpinningReseves.pdf. Last accessed November 2, 2021.

CERC (2017) DOP on Reserve Shutdown and Compensation Mechanism. CERC, New Delhi. https://cercind.gov.in/2017/regulation/SOR132.pdf.

CERC (2020) Report on Short-term Power Market in India: 2019-20. Central Electricity Regulatory Commission, India. https://cercind.gov.in/2020/market_monitoring/Annual%20Report%202019-20.pdf.

Chandra A, Pradhan AK, Sinha AS (2016) PMU based real time power system state estimation using ePHASORsim. *National Power Systems Conference (NPSC) 2016*. doi: 10.1109/NPSC.2016.7858967.

Cole W, Frazier W, Augustine C (2021) Cost Projections for Utility-Scale Battery Storage: 2021 Update. National Renewable Energy Laboratory (NREL). https://www.nrel.gov/docs/fy21osti/79236.pdf.

CSEP (2018) CSEP Electricity & Carbon Tracker. CSEP, New Delhi. https://carbontracker.in/. Last accessed September 14, 2021.

Deloitte (2019). Assessing the flexibility of coal-fired power plants for the integration of renewable energy in Germany. Report prepared for VDKi, October 2019. https://www2.deloitte.com/content/dam/Deloitte/fr/Documents/financial-advisory/economicadvisory/deloitte_assessing-flexibility-coal-fired-power-plants.pdf.

Frick, N M *et al.* (2019) Peak Demand Impacts from Electricity Efficiency Programs. Lawrence Berkeley National Laboratory. Lawrence Berkeley Lab, November 2019. https://eta-publications.lbl.gov/sites/default/files/cost_of_saving_peak_demand_20200902final.pdf.

Gambhir A, Dixit S, Josey A (2020) A critical look at the recent "Round-the-Clock" Supply of 400 MW RE Power tender by SECI. Prayas (Energy Group). https://www.prayaspune.org/peg/resources/power-perspective-portal/242-re-rtc.html.

Helistö et al. (2019) Including operational aspects in the planning of power systems with large amounts of variable generation: A review of modelling approaches. *WIREs Energy Environment*. https://doi.org/10.1002/wene.341.

Hobbs BF (1995) Optimization methods for electric utility resource planning. *Eur J Oper Res*. pp. 1-20. https://doi.org/10.1016/0377-2217(94)00190-N.

Jones-Albergus J (2017) Confronting the Duck Curve: How to Address Over-Generation of Solar Energy. US Dept. of Energy. www.energy.gov/eere/articles/confronting-duck-curve-how-address-over-generation-solar-energy. Last accessed Dec 5, 2021.

Lopes F, Coelho H (2018) Electricity Markets with Increasing Levels of Renewable Generation: Structure, Operation, Agent-based Simulation, and Emerging Designs. Springer Nature, Switzerland.

Mallapragada D *et al.* (2018) Impact of model resolution on scenario outcomes for electricity sector system expansion. *Energy*. https://doi.org/10.1016/j.energy.2018.08.015.

MoEFCC (2015) Thermal plant gazette. Ministry of Environment, Forests and Climate Change, Government of India, Gazette No. 2620, December 8, 2015. https://moef.gov.in/wp-content/uploads/2017/08/Thermal_plant_gazette_scan.pdf.

Ministry of Power (2005) National Electricity Policy. Government of India. https://powermin.gov.in/en/content/national-electricity-policy. Accessed November 15, 2021.

Ministry of Power (2021) MERIT. Government of India. http://meritindia.in/. Last accessed November 2, 2021

NREL (2021). PVWatts Calculator. U.S. Department of Energy. https://pvwatts.nrel.gov/. Last accessed July 27, 2021.

Palchak D, Cochran J (2017) Greening the Grid: Pathways to Integrate 175 Gigawatts of Renewable Energy into India's Electric Grid, Vol. I—National Study. NREL. https://www.nrel.gov/docs/fy17osti/68530.pdf.

PIB (2021a) Switch to solar,will be cost-effective : Gadkari to MSMEs. https://pib.gov.in/PressReleseDetail.aspx?PRID=1702057, accessed November 27, 2021.

PIB (2021b) National Statement by Prime Minister Shri Narendra Modi at COP26 Summit in Glasgow. November 1, 2021. https://pib.gov.in/PressReleasePage.aspx?PRID=1768712. Accessed November 10, 2021.

Sengupta S (2021) Schenley Park to C.R. Park: Exploring Air Pollution, Energy, and Climate in Pittsburgh and New Delhi. Ph.D. Dissertation, Dept. of Engineering & Public Policy, Carnegie Mellon University.

Singh D and Tongia R (2021) Need for an integrated approach for coal power plants, CSEP Discussion Note-02, January 2021. https://csep.org/discussion-note/need-for-an-integrated-approach-for-coal-power-plants.

Srinivasan S., Roshna N *et al.* (2018) Benefit Cost Analysis of Emission Standards for Coal-based Thermal Power Plants in India. (CSTEP-Report-2018-06). https://cstep.in/drupal/sites/default/files/2020-06/CSTEP_RR_BCA_of_Emission_standards_for_TPPs_July2018.pdf.

Staffell I, Pfenninger S (2016) Using bias-corrected reanalysis to simulate current and future wind power output. *Energy* Vol 114. https://doi.org/10.1016/j.energy.2016.08.068.

Tongia R, Harish S, Walawalkar R (2018) Integrating Renewable Energy Into India's Grid −Harder Than It Looks. Brookings India IMPACT Series No. 112018-0. https://www.brookings.edu/wp-content/uploads/2018/11/Complexities-of-Integrating-RE-into-Indias-grid2.pdf.

Tongia R, Sehgal A (eds) (2020) *Future of Coal in India: Smooth Transition or Bumpy Road Ahead?* Notion Press and Brookings India, ISBN: 978-1648288456.

Tongia R (2021) Gas for the Power Sector: Fundamentals Suggest a Niche Role, Chapter in *The Next Stop: Natural Gas and India's Journey to a Clean Energy Future*, V. S. Mehta (ed.), HarperCollins, 2021. ISBN: 978-9390327430.

US National Academy of Engineering (2021) Greatest Achievements of the 20[th] Century. http://www.greatachievements.org/. Last accessed November 30, 2021.

# 9.  Acknowledgements

The author also thanks a number of experts and colleagues who have provided feedback, data, etc. including but not limited to (in alphabetical order) Ashwin Gambhir, Balawant Joshi, Daljit Singh, Danwant Narayanaswamy, Karthik Ganesan, Kaveri Iychettira, Mohua Mukherjee, Nikit Abhyankar, Rahul Walawalkar, Rishabh Jain, Shayak Sengupta, Srihari Dukkipati, Sushil Soonee, Thomas Spencer, as well as several participants who attended CSEP seminars on the research findings. Aarushi Dave provided extensive data gathering and editorial support, and M. Tabish Parray helped with data clean-up and RE analysis. Utkarsh Dalal helped create carbontracker.in, CSEP's tool used for underlying data in the analysis.

# 10. Funding

This research is part of a larger program on energy studies at CSEP. The initial model was funded through a grant by the Shakti Sustainable Energy Foundation and ongoing enhancements are part of studies funded by the MacArthur Foundation.

# 11. Ethics declarations

**Conflict of interest**

The author declares that he has no known competing financial interests in this paper.

**Availability of data and materials**

Data can be shared on request; underlying data available at https://carbontracker.in

**Ethics approval**

Not applicable

**Consent to participate**

Not applicable

**Consent for publication**

Yes.

# Supplementary Information (SI)

## SI 1.  Model Assumptions

**Table 5 Key input assumptions and parametric ranges**

| | | |
|---|---|---|
| Demand Growth Rate | [5%, 5.25%, 5.5%] | |
| Thermal Power Plant Part Load (flexible operations) limit | [55%, 60%, 70%] | |
| Planned RE capacity (Variable RE, without storage) | [250, 300, 350, 400, 450, 500, 550] | GW |
| Solar share within solar + wind | [6/12, 7/12, 8/12] | Base matches CEA (2020)'s optimal fuel mix analysis |
| Prospective Solar CUF | 27% | |
| Prospective Wind CUF | 35% | |
| kWh per kW of solar per day | 5.85 | |
| Solar price in 2021[a] | 43,000,000 | Rs./MW |
| Solar price change annually (nominal) | [-2%, -1%, 0%, 1%] | |
| Wind price in 2021 [a] | 75,000,000 | Rs./MW |
| Wind price in 2030 | 70,500,000 | Rs./MW |
| Foreign Exchange escalation rate | [3%,4%,5%] | |
| Battery Lifespan | 15 (considering upto 5,000 cycles technology) | years |
| Battery depth of discharge min. buffer | 5% | |
| Battery efficiency round trip | 90% | |
| Solar O&M (operations & maintenance) | 600,000 | Rs./MW/year |
| Wind O&M | 500,000 | Rs./MW/year |
| battery and RE O&M inflation rate | 4% | |
| Discount Rate (for NPV calculations) | [6%,8%,10%] | |
| Battery Learning Curve rate in US$ (decline) [b] | [5%,7%,9%,11%] | annually |
| Battery price in 2021 (excludes inverter/Balance of Systems) | 175 | $/kWh |
| Battery price in 2030 | 91.1 [e] | $/kWh (calculated value) |
| INR per USD 2021 | 73.65 | |
| Inverter plus balance of systems cost [c] | 7500 | Rs./kW |
| FGD penalty on coal output | 2.50% | relative basis |
| Auxiliary Consumption | | |
| Thermal (base, pre FGD) | 8% | |
| Gas | 5% | |
| Hydropower | 1% | |
| Nuclear | 7% | |
| RE | 0% | |
| Net retirement of present coal capacity by 2030 | [10, 20] | GW |
| ISTS losses | 3.39% | Estimated from CEA data |

| Fuel costs 2021 - 2019 utilised coal | 2.6 [d] | Rs./kWh |
|---|---|---|
| Fuel costs 2021 - slack coal | see SI Table 6 | |
| Fuel costs 2021 - 2019 utilised gas | see SI Table 6 | |
| Fuel costs 2021 - slack gas | see SI Table 6 | |
| Fuel escalation rate | see SI Table 7 | |
| WACC (weighted avg. cost of capital) | [7.5%, 8.5%, 9.5%] | |
| Grid capacity buffer requirement | 5% | |
| Other RE (non-wind/solar) PLF; ~no growth | 19.80% | |

\* The underlined values in a range are the base values used unless specified otherwise.

[a] Chosen RE capital costs correspond to 2021 costs (LCOE) close to recent bids (slightly higher for reasons given in the text).

[b] The historical fall in battery prices has been higher than 5% annually (BNEF 2021), but this is conservative, including imports and/or any premium for making in India, and also reflects saturation effects over time.

[c] The inverter+balance of systems price has been chosen aggressively low per kW, but we then don't assume any learning curve improvements, instead to maintain quality and have the long lifespan assumed. These prices are only reflective in the second half of the decade (no storage required for VRE in the short run).

[d] There is no official data on average fuel prices; MERIT data (Ministry of Power 2021) are incomplete even if we knew the heat rates (efficiency). Other data sources exclude merchant power, e-auction purchased power, etc. The figure used is an estimate that reflects NTPC plants having a lower variable cost, but distant plants would be more expensive. However, the latter have a lower PLF. The combination of base plus growth reflects costs of FGD upgrades converted into fuel costs, at approximately 0.30 Rs./kWh (average).[30]

[e] While this is higher than some publications like Abhyankar (2021), this is not very different especially after accounting for the low cost for balance of system assumed. In addition, some differential can be covered via import logistics and/or duties.

**Table 6 Key assumptions for new capacity (aka "new something")**

| | Lifespan | Capex [a] Base (2021) [Rs./MW | Capex Escalation Rate per annum | Auxiliary Consumption | Base (2021) Fuel costs (Rs./kWh) | O&M rate (annual, w.r.t. capex) [e] |
|---|---|---|---|---|---|---|
| Coal | 25 | 85,000,000 | 6.0% | 8.0% | 2.4 | 1.5% |
| OCGT | 25 | 50,000,000 | 4.0% | 2.5% | 6.8 | 1.5% |
| CCGT | 25 | 60,000,000 | 5.0% | 5.0% | 5 [b] | 1.5% |
| Gas IC | 18 | 55,000,000 | 4.0% | 0.5% | 5.8 | 1.5% |
| Diesel Gen | 15 | 20,000,000 | 4.0% | 0.5% | 20 [c] | 1.5% |
| Battery w/RE | 13 (~5,000 cycles) | see SI.1 | see SI.1 | 5.0% | Solar as required [d] | 1.5% |
| Inverter (used with Battery) | 13 | 7,500,000 | 0 [f] | Accounted in battery Aux. | n. a. | 1.5% |

[30] This increase in costs for FGD is slightly lower than estimates by Srinivasan *et al.* (2018), but they found costs varying heavily per kWh based on expected output and lifespan. Discussions with technology providers suggest this is a fair estimate for practical costs once competition starts lowering costs of FGD deployment, especially for the majority of coal plants (i.e., excluding outliers or ones slated for retirement soon). This relatively low average also reflects differential PLFs and a weighted average at a system level.

[a] Author estimates based on discussions with domain experts and literature; chose to allow a representative LCOE based on fuel costs.

[b] This translates to $8/MMBTU natural gas costs, with a high ramping duty cycle (thus, lower efficiency than theoretical). Gas estimates are from Tongia (2021), driven by imported LNG.

[c] This price is assumed for biodiesel as well. Future prices for biodiesel are thus very high, one reason under-sizing a battery isn't cost-effective. There is no reason that with innovation and R&D, e.g., third-generation biofuels, the price may not be as high as conservatively projected.

[d] While batteries have no fuel, they require charging, which is assumed from solar, which entails capital costs, except to the extent surplus or curtailed RE is utilised. We purposely chose to avoid coal-based battery for our base analysis. Allowing such battery charging could reduce RE for battery requirements, but also raise emissions.

[e] Regulators often cap O&M charges; these are representative figures for future builds. In practice OCGT O&M will be lower than CCGT, but the capital costs are also lower.

[f] Inverter capex is assumed constant because the base price is aggressive, and there are needs for quality improvements compared to today's deployments; in rupee terms we already see hardening of prices today.

## SI 2.  Methodology for calculating battery displacement of fossil fuels

Once the model produces daily surplus battery capability (based on unmet aka NEW demand, as well as charging to the battery via curtailed RE and dedicated solar), this surplus can then displace gas and coal. By having separate prices for natural gas utilized in 2019 (assumed to be less expensive) and subsequent gas (labelled non-APM), we can prioritize battery surplus output despatch for non-APM gas before coal.[31]

For displacing coal, in addition to simply displacing coal like gas, there is another benefit based on lowering the daily maximum, which reduces curtailment due to flex operations limits for coal (nominally 60% of daily high on average). This further reduces future curtailment as well, a sort of multiplier effect. In fact, based on the profile of fuel-wise output, a 10% reduction in coal can have more than just 60% times 10% reduction in curtailment, since one MW of coal reduction in a day can apply to many hours of RE curtailment in the same day.

In the model, each day's coal output is stacked to create a declining curve. Going down the curve, the area under the curve (kWh by declining coal levels) is matched with surplus battery energy (kWh) to determine the decline in daily maximum coal capacity utilized (MW). This is then multiplied by the coal flex limit to determine a new daily limit for RE output, which then calculates a bonus reduction in RE curtailment across each time period. SI Fig. 17 shows that this saving can be substantial and non-linear, more so on a percentage basis. For lower capacities of battery chosen (not shown), the relative benefit is higher. The reason for the upswing seen in the figure is increasing peakiness of unmet (NEW) demand over time.

**Fig. 17 Additional RE curtailment avoided (and hence savings on coal fuel) by lowering daily maximum coal**

[31] This same differential is used for valuing NEW coal capacity which could be surplus compared to NEW demand.

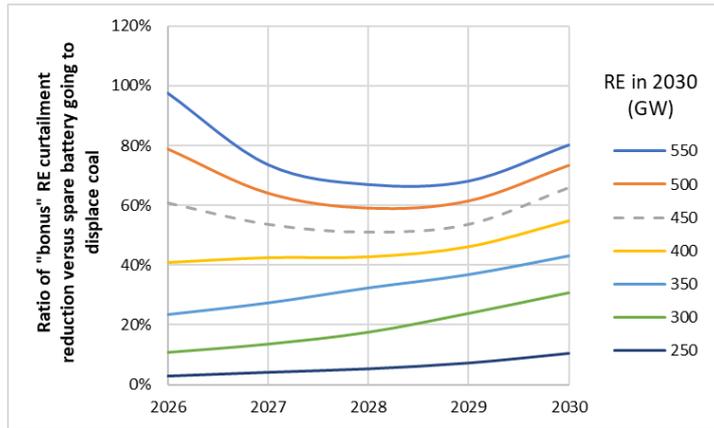

\* 2025, the first year with NEW supply, isn't shown because it's an outlier with very little volume (hence, high ratios).

Examining batteries in detail in 2030 and displacement effects, in absolute terms for the base case of 450 GW RE and 2:1 solar:wind, RE curtailment becomes 276.8 TWh and the spare battery annually is 170.1 TWh (with a full sizing of the battery and no additional voluntary solar added beyond that necessary for charging the battery to meet NEW demand). Out of this curtailed RE, 53.8 TWh displaces non-APM gas (based on its expensive nature), followed by 116.3 TWh displacing coal. However, this is before the bonus created by lowering the daily peak of coal, which then leads to lower RE curtailment (and thus further fuel savings of coal) of 66.9 TWh.

## SI 3.  SI: Fossil fuel price escalation over time and implications

**Table 7 Fossil Fuel escalation rates (applicable for existing fossil capacity and NEW supply)**

| | |
|---|---|
| **Coal\*** | 5% |
| **Natural Gas\*\*** | 3% |
| **Diesel / Biodiesel\*\*** | 3% |

*\* Historical coal price rises have been disproportionally due to levies (like the "coal cess") or freight charges, and less driven by mining costs (Tongia and Sehgal, 2020). The base rise in coal prices is chosen to be close to historical levels and also inflation, which has been falling over time, but there is great uncertainty over a nine-year time period. This is especially true when we consider landed cost including transportation. A key factor for potential rise in prices would be a carbon tax (or increase in "coal cess"), there are also potential factors to keep per unit (kWh) price rises in check, including increase in deployment and utilization of more efficient coal power plants, reduction in coal transportation costs through a combination of more local delivery and/or deployment of dedicated freight corridors for railways, and the rise of captive and private mining, that too in favourable locations (near power plants and/or with superior geo-technicals like stripping ratio, i.e., minimal overburden removal).*

*\*\* Includes regular diesel and biodiesel. While diesel prices have risen, there have been disproportional rise in taxes, and not crude oil prices (at time of analysis). The baseline prices are high, post-covid with high taxes, and hence future growth is assumed moderate. This would also be the case if petroleum products become part of the GST regime. The focus in this paper for NEW supply is biodiesel, esp. to offset capital investments in NEW.*

One key reason for the superiority of a battery for NEW supply is the high price for fossil fuels on a nominal basis thanks to inflation. By 2030, the nominal fuel cost of coal generation is 4.03 Rs./kWh, while solar falls below 2 Rs./kWh on a nominal basis, despite rising O&M costs. The base fall in solar costs per MW construction are 2% per annum.

This means a battery has high value when it offsets expensive non-APM gas first, and then coal, which blended together typically exceed the 2030 loaded capital costs of a battery (see SI Fig. 18). Otherwise, the duty cycle issues as shown in Fig. 16 means a lower value for higher battery sizing.

**Fig. 18 Base fossil fuel prices over time**

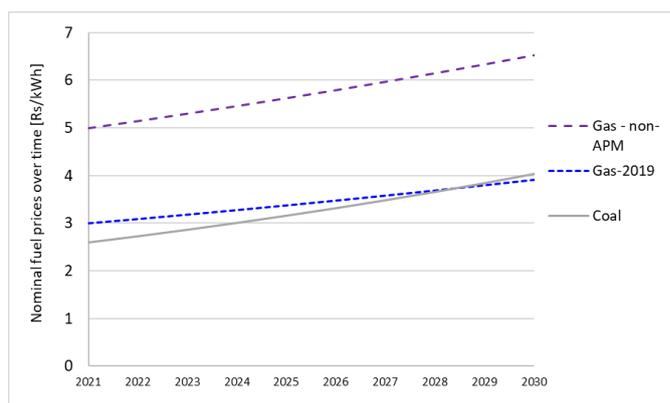

*\* APM gas is cheaper and the majority if not close to all the use in 2019. Coal is cheaper on average in the near term, but its price may rise faster over time for various reasons, including environment concerns, and the fact that APM gas is prioritized. Policy changes in fuel pricing would, naturally, shift these numbers significantly, including the shift to GST for gas.*

It is very difficult to compute the practical costs of a battery per kilowatt-hour as the costs depend on the duty cycle as well as cost of RE to charge the battery (e.g., free RE from curtailment). If, hypothetically, one had to pay for RE (solar) to charge the battery, and its value was independent of the value of the fuel it would displace, e.g., expensive non-APM gas, then fully loaded costs for a battery including RE could well exceed 6 Rs./kWh in 2030, even with a 7% annual decline in battery capex costs (in US$ terms).

## SI 4.  Under-sizing peak capacity for new – balanced by (say) biodiesel

**Fig. 19 Secondary unmet demand share in 2030 by under-sizing the battery (potentially met via biodiesel)**

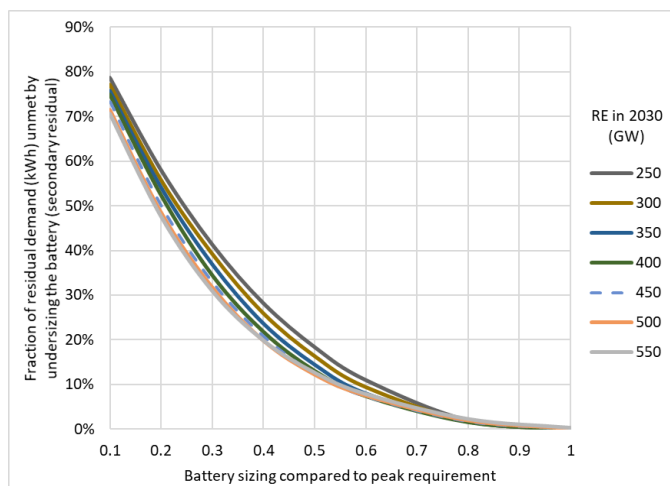

*\* It assumes full daily charging of chosen battery capacity, and thus sufficient RE (regardless of whether based on curtailed RE or dedicated solar).*

The theory for under-sizing a battery has already been covered in the main text. One issue is suitable availability of a low-carbon replacement.

Use of biofuels at a commercial scale has a mixed track record in India at best. Most applications focused on displacing liquid fuels for transportation with bioethanol and biodiesel, where the benchmark has been a much higher dollars per MMBtu than comparison with fuels for the power sector such as coal. In this analysis, we apply a high biodiesel cost per kilowatt hour, based on a reference price from traditional diesel. We assume Rs. 20/kWh as the variable cost in 2021.

The total system savings by under-sizing are less than SI Fig. 19 would suggest because of the need to add capex of biodiesel generation (which costs more per kW than an inverter). Even though we add capacity from biodiesel, lower inverter capacity requirements are restricted to match worst-case in the year instantaneous battery supply.[32] One other finding is that outlier time periods influence a significant fraction of systems costs.

Because of the non-linearity in under-sizing the energy of the battery, the normalised secondary unmet demand by total annual energy requirements (in kWh) shows it is usually a fraction of a percent of total demand for high RE scenarios (SI Fig. 20). This is assuming full daily charging of the battery, our base case.

**Fig. 20 Secondary residual unmet demand compared to total annual demand in 2030**

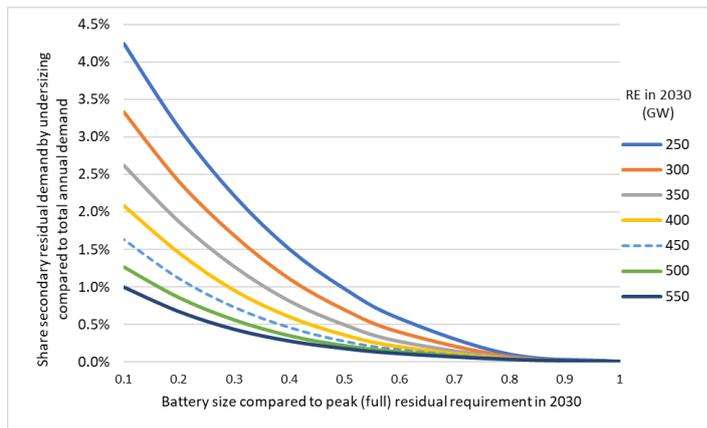

* It assumes sufficient secondary (solar) to charge the battery.

Because of the strong nonlinearity between battery (or coal) capacity savings and consequent unmet secondary residual demand needing energy (such as from biodiesel), the volume of biodiesel required is modest. One litre in a large, efficient generator can produce at least 4 kWh of electricity. In terms of litres and land requirements, assuming 850 litres jatropha production per hectare, and 4.3 kWh/litre, 10 billion kWh per year would need 27,360 sq. km. of land. That is a quarter of the requirement for the proposed 20% ethanol blending target of the government.

With battery under-sizing down to 40% of the peak and 450 GW of RE, the annual energy required for secondary residual unmet demand is under 5 TWh (5 billion units), or half the above. The total volume is still substantial but at a national scale this is still likely achievable.[33] An alternative worth exploring is the use of demand response to shift load. As of now, the demand response market in India is nascent.[34]

Batteries are optimally sized as full due to the value of displacing expensive fossil fuels due to spare capability (beyond meeting NEW supply). Counter-intuitively, lower VRE installed leads to more spare battery capability. While lower VRE reduces curtailed RE to charge the battery, the larger size of battery required for NEW demand means more harnessing of whatever curtailed RE there is, combined with the relatively low cost of adding solar to charge the battery. This is an example of many non-linearities in optimal sizing.

## SI 5.  SI: Additional solar for charging the battery.

One variable in the model is the choice of how much spare solar to build for the battery. The minimum (and mandatory) level is that required to meet NEW demand (normalized as zero voluntary dedicated solar), but one can choose to add more solar to harness surplus battery capacity up to a daily full charge (normalized as sizing of 1 for the solar). However, this value depends significantly on VRE levels of deployment, which determine both surplus curtailment as well as surplus battery capacity that awaits charging. SI Fig. 21 shows the solar capacity required beyond the VRE planned as per targets for different levels of RE in 2030 and battery sizing. The requirements are substantial, much more than the inverter capacity (~peak NEW demand in MW). This is because some days require long-duration battery output, and a single kW of dedicated solar can only power a little under 6 kWh of battery output on average (inclusive of battery efficiency losses).

---

[32] There is insufficient data on battery capabilities for rapid charging and discharging. We assume 1C capabilities at best, meaning a 1 kWh-hour battery will take one hour to charge (or slightly longer inclusive of efficiency penalties).

[33] Plants inherently only convert ~1-2% of incoming solar energy into biomass, and conversion to specific fuels like biodiesel further reduces the energy harvest. This is one reason that per square kilometre, solar offers 1-2 orders of magnitude greater electricity. Palm oil production has a higher yield per square kilometre than jatropha, but has much greater water requirements and other land implications.
[34] India has limited retail Time of Day pricing for consumers, primarily available (or, in a few cases like New Delhi, mandatory) for bulk consumers.

**Fig. 21 Solar required in 2030 to charge the battery to meet NEW demand or pro-rata**

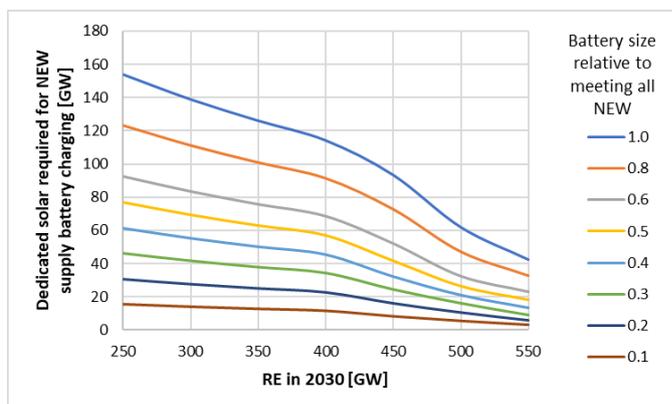

* This is the solar required for meeting NEW supply, and excludes additional solar to harness the battery in full. This is for various sizes of battery chosen compared to meeting full unmet demand (NEW supply).

## SI 6.   SI: Economic dispatch generation mix and coal output over time

A key question is whether India will achieve a peak in coal output. Coal's required output varies over time (Fig. 12), and one key factor is the growth of RE (SI Fig. 22, shown for 2030).

**Fig. 22 Generation Mix for varying RE in 2030 (a) prior to NEW; (b) with battery as NEW**

(a)                                                                                  (b)

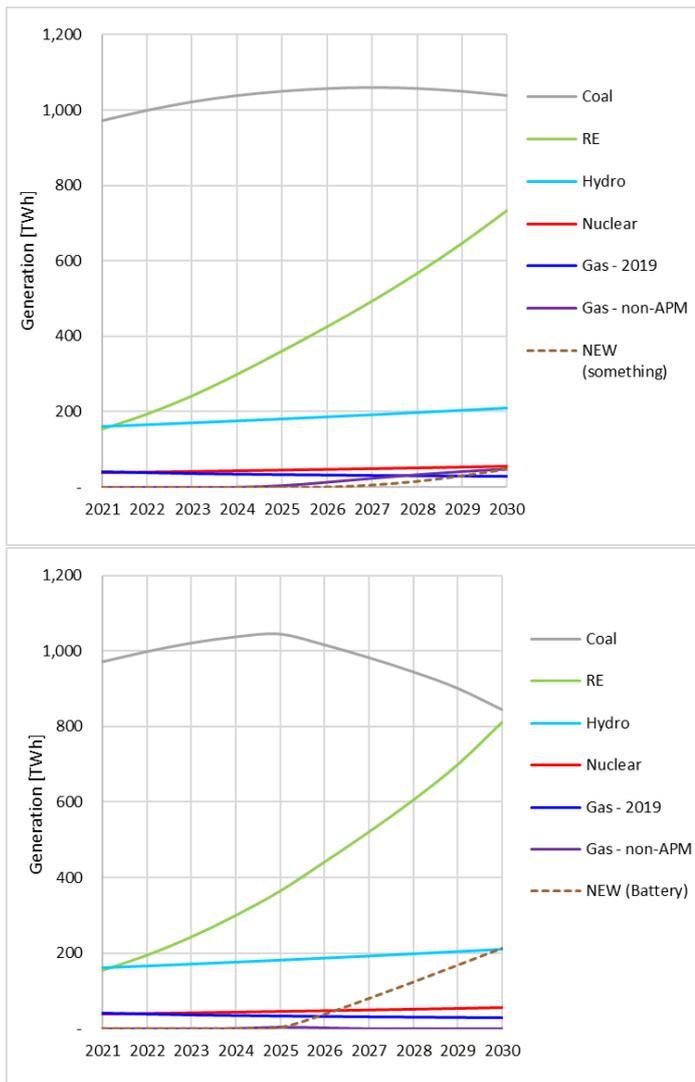

*This is for 5.25% demand growth, 2:1 solar:wind, and full battery sizing for NEW supply.*

*\** *RE going through the battery is not shown under RE, but is part of NEW (battery).*

*\*\** *A battery can displace both gas and coal up to a point. The total demand in 2030 is 2160 TWh.*

The total output of coal with and without NEW (as a battery) is shown in SI Fig. 23.

**Fig. 23 Coal output over time (a) before NEW; (b) with batteries as NEW**

(a)                                                                                                    (b)

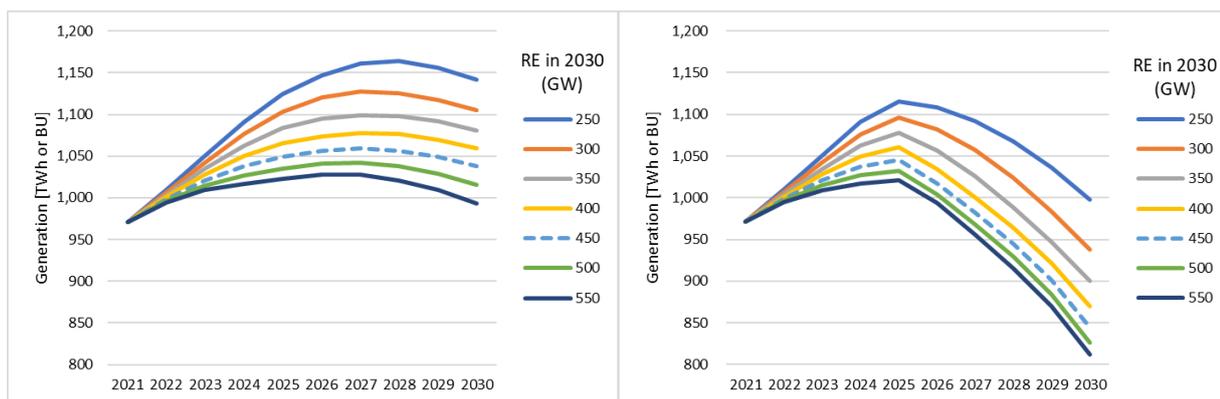

Surprisingly, there is "peak coal" between 2027-28 even before considering batteries to supply NEW. This is not a prediction of optimal, but a recognition that unless more coal capacity is added (and we assume some capacity retirement), it reaches its limits of supply matching demand, especially factoring in time of day. Under-construction coal plants if built will shift this slightly.

Coal fleet PLF is an important check of performance, reflective of not only the output but also capacity (SI Fig. 24). We see that it increases and then declines in the pre-NEW case, and declines more sharply with battery deployments. The former was growth up to likely fleet limits, while the latter is manageable and also in line with reduced PLFs anticipated by CEA (2020) in a high RE scenario.

### Fig. 24 Coal PLF (a) before NEW supply; (b) with batteries as NEW

(a)                                                                                                    (b)

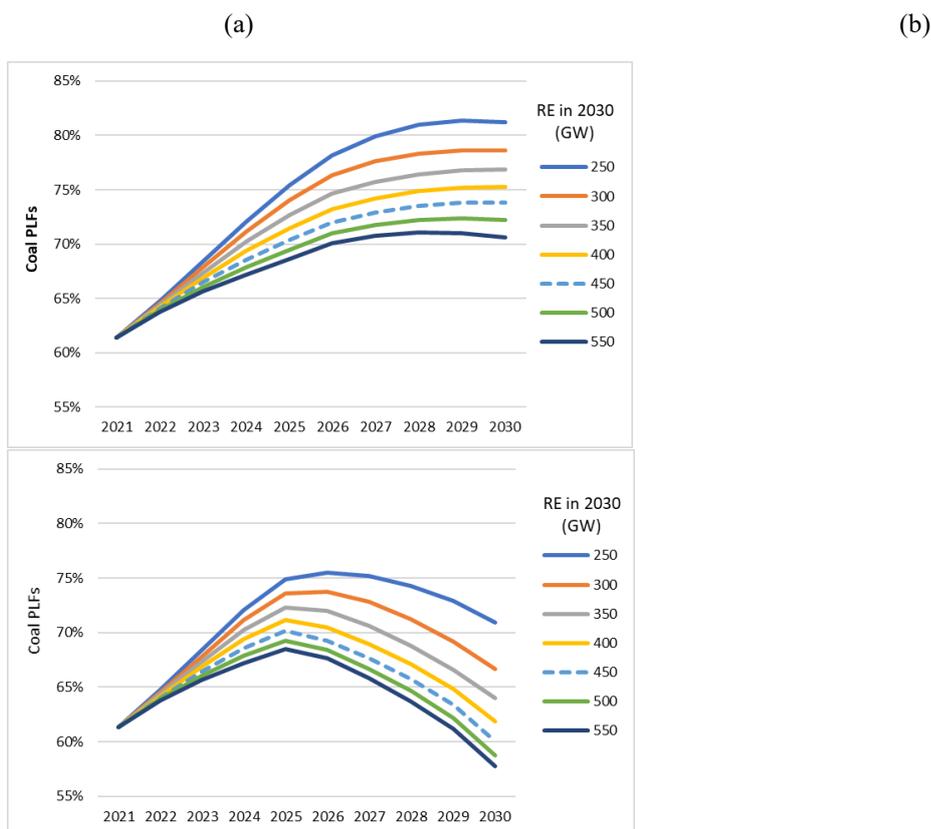

*\* The rising thermal PLF isn't only rising generation but an assumption of limited retirement of existing capacity by 2030. Coal is a backstop generator, and hence at lower RE deployments its output grows measurably, to the level approaching the practical limits.*

A key component of lower coal output is rising RE, not merely from chosen installed base by 2030 but also the expectation of higher-than-historical PLF of RE from new builds. Lower CUFs or PLFs of prospective RE would delay peak coal, but it would still happen within 1-2 years (comparing growths of output across coal vs.

RE). This is because the impact is not as much as a simplistic percentage reduction in RE output would indicate, say 3% in absolute terms. First, not all RE by 2030 is future RE, with possibilities for less-than-modelled PLF. Second, there is measurable curtailment, and lower RE output lowers curtailment as well. Lastly, the uncertainty in output is likely higher for wind than solar, which has a much lower share of RE *growth* through 2030. On the other hand, achieving less than even 2:1 solar:wind would reduce the future RE output.